\documentclass[11pt,a4paper]{article}

\usepackage{mathtools}
\usepackage{authblk} 
\usepackage{algpseudocode,algorithmicx,algorithm}

\usepackage{mathrsfs}
\usepackage{latexsym,bm}
\usepackage{amsmath,amsfonts,amssymb,amsthm}
\usepackage{extarrows}

\usepackage{graphicx,subfigure,epstopdf,float}
\usepackage{enumerate,cases,multirow}
\usepackage{makecell}
\usepackage{caption}

\usepackage{longtable,colortbl,arydshln,threeparttable}
\definecolor{mygray}{gray}{.9}

\usepackage{indentfirst}
\usepackage[top=25mm,bottom=20mm,left=25mm,right=20mm]{geometry}
\usepackage{cite}

\usepackage{listings}

\usepackage{makeidx}        
\usepackage{booktabs}
\usepackage[bookmarksnumbered,colorlinks,citecolor=red,linkcolor=red,hyperindex,linktocpage=true]{hyperref}

\newcommand{\ket}[1]{| #1 \rangle} 
\newcommand{\bra}[1]{\langle #1 |} 
\newcommand{\braket}[2]{ \langle  #1 | #2  \rangle }

\newcommand{\bb}{\boldsymbol}

\def \d {{\rm{d}}}
\def \e {{\rm{e}}}
\def \i {{\rm{i}}}

\newcounter{parentalgorithm}

\makeatother

\newtheorem{theorem}{Theorem}[section]
\newtheorem{lemma}{Lemma}[section]

\theoremstyle{remark}
\newtheorem{remark}{\bf Remark}[section]

\numberwithin{equation}{section}

\usepackage{physics}

\begin{document}

\title{Quantum simulation of the Fokker-Planck equation via Schr\"odingerization}

\author[1]{Shi Jin\thanks{shijin-m@sjtu.edu.cn}}
\author[1, 2]{Nana Liu\thanks{nana.liu@quantumlah.org}}
\author[3]{Yue Yu\thanks{terenceyuyue@xtu.edu.cn}}
\affil[1]{School of Mathematical Sciences, Institute of Natural Sciences, MOE-LSC, Shanghai Jiao Tong University, Shanghai, 200240, China}
\affil[2]{University of Michigan-Shanghai Jiao Tong University Joint Institute, Shanghai 200240, China}
\affil[3]{School of Mathematics and Computational Science, Hunan Key Laboratory for Computation and Simulation in Science and Engineering, Key Laboratory of Intelligent Computing and Information Processing of Ministry of Education, Xiangtan University, Xiangtan, Hunan 411105, PR China}

\maketitle

\begin{abstract}
    This paper studies a   quantum simulation technique for solving the Fokker-Planck equation.
    Traditional semi-discretization methods often fail to preserve the underlying Hamiltonian dynamics and may even modify the Hamiltonian structure, particularly when incorporating boundary conditions.
    We address this challenge by employing the Schr\"odingerization method --  it converts any linear partial and ordinary differential equation with non-Hermitian dynamics into systems of Schr\"odinger-type equations. It does so via the so-called warped phase transformation that maps the equation into one higher dimension.
     We explore the application in two distinct forms of the Fokker-Planck equation.
     For the conservation form, we show that the semi-discretization-based Schr\"odingerization is preferable, especially when dealing with non-periodic boundary conditions.
    Additionally, we analyze the Schr\"odingerization  approach for unstable systems that possess positive eigenvalues in the real part of the coefficient matrix or differential operator. Our analysis reveals that the direct use of Schr\"odingerization has the same effect as a stabilization procedure.
  For the heat equation form, we propose a quantum simulation procedure based on the time-splitting technique, and give explicitly  its corresponding quantum circuit.
  We discuss the relationship between operator splitting in the Schr\"odingerization method and its application directly to the original problem, illustrating how the Schr\"odingerization method accurately reproduces the time-splitting solutions at each step.
  Furthermore, we explore finite difference discretizations of the heat equation form using shift operators.
  Utilizing Fourier bases, we diagonalize the shift operators, enabling efficient simulation in the frequency space.
Providing additional guidance on implementing the diagonal unitary operators, we conduct a comparative analysis between diagonalizations in the Bell and the Fourier bases, and show   that the former generally exhibits greater efficiency than the latter.
\end{abstract}


\tableofcontents

\section{Introduction}

The Fokker-Planck equation is a fundamental evolution equation used to model various stochastic processes in statistical mechanics, plasma physics, fluid dynamics, neural networks, machine learning, and socio-economic modeling \cite{Pav,McKean,Risken1989,Hu2023PreservingFP,Escande07FP,CACERES20111084,Tabandeh22physics,Moroni06FP,Jin2024VFPL,Jin2018VPFP}, among others. It governs the behavior of a wide class of Markov processes and is particularly important for describing systems influenced by drag forces and random forces. The linear form of the Fokker-Planck equation is given by
\begin{equation}\label{FP0}
    \partial_t f = \nabla \cdot (f \nabla V(x)) + \sigma \Delta f,
\end{equation}
In this equation, $f = f(t,x) \geq 0$ represents the unknown density function, $V(x)$ is an external potential, and $\sigma > 0$ is a constant. The equation describes the time evolution of the probability density function of a particle's position. The first term on the right-hand side is the drift term, while the second term represents diffusion generated by white noise.
It is worth noting that the Fokker-Planck equation possesses a steady-state solution given by $f=\exp(-V(x)/\sigma)$ \cite{Pav}. In this context, we assume periodic boundary conditions, where $x =(x_1,\cdots,x_d)\in \Omega = (-1,1)^d$, for convenience.

A main challenge in classical numerical simulation of  the Fokker-Planck equation is the curse-of-dimensionality, since it is typically high-dimensional partial differential equation (PDE) in statistical physics and machine learning, for example.
In recent years there is increasing activity in developing quantum algorithms to be used on quantum computers -- yet to be developed in the future -- to solve PDEs \cite{Cao2013Poisson,Berry-2014,qFEM-2016,Costa2019Wave,Engel2019qVlasov,Childs-Liu-2020,Linden2020heat,
Childs2021high,JinLiu2022nonlinear,GJL2022QuantumUQ,JLY2022multiscale}, many of which rely upon the exponential speedup advantages in quantum linear systems of equations \cite{HHL2009,Childs2017QLSA,Costa2021QLSA,Berry-2014,BerryChilds2017ODE,Childs-Liu-2020,Subasi2019AQC}.
A strategy for crafting quantum PDE solvers involves discretizing spatial variables to yield a system of ordinary differential equations (ODEs), which can then be tackled using quantum ODE solvers \cite{Berry-2014,BerryChilds2017ODE,Childs-Liu-2020}. Notably, if the solution operator to the resulting ODE system is a unitary system,   quantum simulations can be  executed with reduced time complexity compared to quantum ODE solvers or other quantum linear algebra solvers (like quantum difference methods) \cite{Berry-2014,JLY2022multiscale}. If the system is not a unitary one needs to ``dilate'' it to a unitary system
\cite{JLY22SchrLong,Javier2022optionprice,Burgarth23dilations,CJL23TimeSchr}.

In a recent work, a new, simple and generic framework coined {\it Schr\"odingerization} was introduced in \cite{JLY22SchrShort, JLY22SchrLong} that allows quantum simulation for {\it all} linear PDEs and ODEs. The idea is to use a warped phase transform that maps the equations to one higher dimension, which, in the Fourier space, become a system of Schr\"odinger type equations!
The approach has been expanded to tackle various problems, including open quantum systems in a bounded domain where artificial boundary conditions -- which are not unitary operators -- are needed \cite{JLLY23ABC}, problems entailing physical boundary or interface conditions \cite{JLLY2024boundary}, Maxwell's equations \cite{JLC23Maxwell}, ill-posed scenarios such as the backward heat equation \cite{JLM24SchrBackward}, linear dynamical systems with inhomogeneous terms \cite{JLM24SchrInhom}, non-autonomous ordinary and partial differential equations \cite{CJL23TimeSchr}, and iterative linear algebra solvers \cite{JinLiu-LA}.

 Among the PDE examples Schr\"odingerized in \cite{JLY22SchrLong} is the Fokker-Planck equation.  Two formulations were used there:  the conservation form and the heat equation forms. This paper studies in detail the corresponding quantum algorithms. We summarize the scope of our study and findings here:
\begin{enumerate}[1)]
  \item For the conservation form, we discuss two treatments to Schr\"odingerize the differential equation: discretize in space first and then Schr\"odingerize, or Schr\"odingerize first and then discretize in space.
      We find that the former is preferable because, even under periodic conditions, the second method doesn't consistently produce a Hamiltonian system when directly performing the discrete Fourier transform on the auxiliary variable $p$, although it can be transformed to the first one easily. It also incurs additional computational overhead for problems with non-periodic boundary conditions.

  \item With the assistance of the findings in \cite{JLM24SchrInhom}, we analyze two Schr\"odingerization-based approaches for unstable systems that have positive eigenvalues in the real part of the coefficient matrix. The first approach involves using the exponential change of variables in \cite{ALL2023LCH} to obtain a stable equation, followed by the application of the Schr\"odingerization technique. The second approach is to directly employ the Schr\"odingerization technique as in \cite{JLM24SchrInhom}.
      Both methods have comparable impact on projecting onto the solution vector.

  \item For the heat equation form, we propose a quantum simulation procedure based on the time-splitting technique, and construct its corresponding quantum circuits. We discuss the relation between operator splitting in the Schr\"odingerization method and its direct application to the original problem, showing how the Schr\"odingerization method accurately reproduces the time-splitting solutions at each step.
   Additionally, we explore finite difference discretizations of the heat equation form by employing shift operators. Utilizing the Fourier basis, we diagonalize these shift operators in the frequency space, thus simplify the problem to the evolution of diagonal unitary operators. Furthermore,   we compare  the two diagonalizations,  in the Bell and Fourier bases respectively, and find that  the diagonalization in the Bell basis generally exhibits greater efficiency than in the Fourier basis.
\end{enumerate}

The paper is structured as follows: In Section \ref{sec:Schr}, we provide an overview of the Schr\"odingerization approach. Section \ref{sec:conservation} delves into the quantum simulation of the conservation form of the Fokker-Planck equation, exploring two different implementations  of the Schr\"odingerization procedure and analyzing the projection probability for the unstable system generated by the spectral discretization of the conservation form. In Section \ref{sec:FKheat} we focus on the quantum simulation procedure for the heat equation form by employing the time-splitting technique. We examine the relationship between operator splitting in the Schr\"odingerization method and the original problem, as well as explore finite discretizations using shift operators. Conclusions are presented in the final section.

\section{Overview of the Schr\"odingerization method} \label{sec:Schr}

\subsection{The Schr\"odingerization method}

Consider the following linear ordinary or partial differential equations:
\begin{equation}\label{ODElinear}
 \begin{cases}
 \frac{\d \bb{u}(t)}{\d t} = A(t) \bb{u}(t) + \bb{b}(t), \\
 \bb{u}(0) = \bb{u}_0,
 \end{cases}
\end{equation}
where $\bb{u}, \bb{b} \in \mathbb{C}^n$  and $A \in \mathbb{C}^{n\times n} $. $A$ is a linear operator (for ODEs) or linear differential operators (for PDEs). In general, $A $ is non-Hermitian, i.e.,  $A^{\dagger} \neq A$, where "$\dagger$" denotes conjugate transpose. Throughout the paper, we assume that $A(t) = A$ is independent of time and $\bb{b}(t) = \bb{0}$ for simplicity. For time-dependent $A$, the so-called non-adiabatic problem, we refer to \cite{CJL23TimeSchr}.

We begin by  decomposing $A$ into a Hermitian term and an anti-Hermitian term:
\[A = H_1 + \i H_2, \qquad \i = \sqrt{-1},\]
where
\[
H_1 = \frac{A+A^{\dagger}}{2} = H_1^{\dagger}, \qquad H_2 = \frac{A-A^{\dagger}}{2 \i} = H_2^{\dagger}.
\]
A natural assumption is that the semi-discrete system \eqref{ODElinear} inherits the stability of the original PDE, which implies that $H_1$ is negative semi-definite. That is, there exists a unitary matrix $Q$ such that $Q^{-1} H_1 Q = \text{diag}(\lambda_1, \cdots, \lambda_n)=:\Lambda$ with $\lambda_j\le 0$.

\begin{figure}[!htb]
  \centering
  \includegraphics[height=5cm,width=10cm]{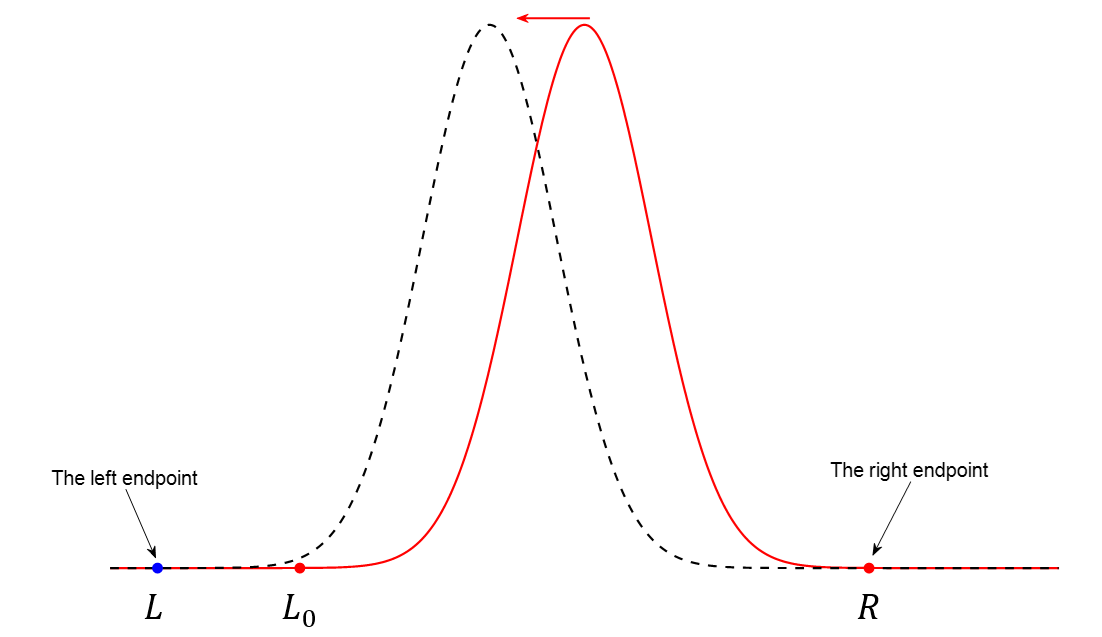}\\
  \caption{Schematic diagram for the computational domain of $p$}\label{fig:domain}
\end{figure}

Using the warped phase transformation $\bb{v}(t,p) = \e^{-p} \bb{u}(t)$ for $p\ge 0$ and symmetrically extending the initial data to $p<0$,  system \eqref{ODElinear} is  then transformed to a system of linear convection equations \cite{JLY22SchrShort,JLY22SchrLong}:
\begin{equation}\label{u2v}
\begin{cases}
 \frac{\d}{\d t} \bb{v}(t,p) = A \bb{v}(t,p) = - H_1 \partial_p \bb{v} + \i H_2 \bb{v}, \\
 \bb{v}(0,p) = \e^{-|p|} \bb{u}_0.
 \end{cases}
\end{equation}
For numerical implementation, it is natural and convenient to introduce  $\alpha = \alpha(p)$ in the initial data of \eqref{u2v} for $p<0$:
\begin{equation}\label{u2valpha}
\begin{cases}
 \frac{\d }{\d  t} \bm{v}(t,p) = A \bm{v}(t,p) = - H_1 \partial_p \bm{v} + \i H_2 \bm{v}, \\
 \bm{v}(0,p) = \e^{-\alpha(p) |p|} \bm{u}_0.
 \end{cases}
\end{equation}
To match the exact solution, $\alpha(p) = 1$ is necessary for the region $p> 0$. In the $p>0$-domain, we will truncate the domain at $p=R$, where $R$ is sufficiently large such that $e^{-R} \approx 0$. We will choose a large $\alpha$ for $p<0$ so the solution (see Fig.~\ref{fig:domain}) will have a support within a relatively small domain.  Let $\tilde{\bb{v}} = Q^{-1} \bb{v}$.  Since  the wave $\tilde{\bm{v}}_j$ moves to the left, we choose an artificial boundary at $p=L<0$, for $|L|$ large enough such that $\tilde{\bm{v}}_j$, initially almost compact at $[L_0, R]$, will not reach the point $p=L$ during the duration of the computation. This will allow us to use a periodic boundary condition in $p$ in the spectral approximation.

We now discuss how to restore the solution $\bm{u}(t)$. A direct strategy, in light of the relation $\bm{u}(t)=\e^p \bm{v}(t,p)$ for all $p>0$, is to simply choose any $p_{k_*}>0$ and let
\begin{equation}\label{v2psipstar}
\bm{u}(t) = \e^{p_{k_*} t} \bm{v}(t,p_{k_*}).
\end{equation}
A more intuitive view is by discretizing the $p$ domain and concatenating the corresponding function for each $p$.  Toward this end,  we choose uniform mesh size $\Delta p = (R-L)/N_p$ for the auxiliary variable with $N_p = 2N$ being an even number, with the grid points denoted by $a = p_0<p_1<\cdots<p_{N_p} = b$.
To compute $\bm v(t,p),$ let the vector $\bm{w}$ be the collection of the function $\bm v$ at these grid points, defined more precisely as follows,
\[\bm{w} = [\bm{w}_1; \bm{w}_2; \cdots; \bm{w}_n], \]
with ``;'' indicating the straightening of $\{\bm{w}_i\}_{i\ge 1}$ into a column vector. This can also be expressed as a superposition state using $\ket{k}$ as a new basis,
\[
 \bm{w}_i = \sum_k \bm{v}_i (t,p_k) \ket{k},
\]
where $\bm{v}_i$ is the $i$-th entry of $\bm{v}$.

By applying the discrete Fourier transformation in the $p$  direction, one arrives at
\begin{equation}\label{heatww}
\frac{\d}{\d t} \bb{w}(t) = -\i ( H_1 \otimes P_\mu ) \bb{w} + \i (H_2 \otimes I) \bb{w}.
\end{equation}
 Here, $P_\mu$ is the matrix expression of the momentum operator $-\i\partial_p$, given by
\[P_\mu = \Phi D_\mu \Phi^{-1},  \qquad D_\mu = \text{diag}(\mu_{-N}, \cdots, \mu_{N-1}), \]
where $\mu_l = 2\pi l/(R-L)$ are the Fourier modes and
\[\Phi = (\phi_{jl})_{M\times M} = (\phi_l(x_j))_{N_p\times N_p}, \qquad \phi_l(x) = \e ^{\i \mu_l (x-L)} \]
is the matrix representation of the discrete Fourier transform.
At this point, we have successfully mapped the dynamics back to a Hamiltonian system.
By a change of variables $\tilde{\bm{w}} = (I \otimes \Phi^{-1})\bm{w}$, one has
\begin{equation}\label{generalSchr}
\frac{\d}{\d t} \tilde{\bb{w}}(t) = -\i ( H_1 \otimes D_\mu ) \tilde{\bb{w}} + \i (H_2 \otimes I) \tilde{\bb{w}}.
\end{equation}
This is more amenable to an approximation by a quantum algorithm. In particular, if $H_1$ and $H_2$ are sparse, then \eqref{generalSchr} is a Schr\"odinger equation with the Hamiltonian $H = H_1 \otimes D_\mu - H_2 \otimes I$ that inherits the sparsity.

\begin{remark} The  initial value of $\bb{w}$ is in $C$ but not $C^1$. Thus even if a spectral method is used in the discrertization of the $p$-derivative, the numerical accuracy in $p$ is only first order. Higher order accuracy in $p$ can be achieved by choosing smoother initial data for $\bb{w}$. See \cite{JLM24SchrInhom,JLM24SchrBackward}.
\end{remark}

\begin{remark}
If one is concerned about both the real and imaginary parts of $\bb{u}$, we can denote $\bb{u} = \bb{u}_R + \i \bb{u}_I$ and define $\widetilde{\bb{u}} = [\bb{u}_R, \bb{u}_I]^T$ to obtain
\[\frac{\d }{\d t} \widetilde{\bb{u}}(t) = (\sigma_x \otimes H_1) \widetilde{\bb{u}}(t)  - \i (\sigma_y \otimes H_2) \widetilde{\bb{u}}(t), \]
where $\sigma_x$ and $\sigma_y$ are $2\times 2$ Hermitian matrices, given by
\[\sigma_x = \begin{bmatrix} 1 & 0 \\ 0 & 1 \end{bmatrix}, \qquad \sigma_y = \begin{bmatrix} 0 & -\i \\ \i & 0 \end{bmatrix}.\]
\end{remark}

In what follows we assume $\alpha(p) \equiv 1$.

\subsection{Retrieval of the solution}

With the state vector encoding $\tilde{\bm{w}}$, one can first apply the quantum Fourier transform on $p$ to get back to $\bm{w}$ in \eqref{heatww}, which can be written as the tensor-product form
  \[\bm{w}(t) = \sum\limits_{i,k} w_{ik}(t) \ket{i}\ket{k}, \qquad w_{ik}(t) \equiv \bm{v}_i(t,p_k),\]
and then restore the solution $\bm{u}$ via the simple relation in \eqref{v2psipstar}.  In fact, if the direct formula \eqref{v2psipstar} is used, one can apply quantum measurement to the $p$-register in the computational basis. The state vector is then collapsed to
  \[ \ket{\bm{\psi}_*} \equiv \frac{1}{\mathcal{N}}\Big(\sum\limits_{i} w_{ik_*} \ket{i} \Big) \otimes \ket{k_*}, \quad
  \mathcal{N} = \Big(\sum\limits_i |w_{ik_*}|^2 \Big)^{1/2}\]
  for some $k_*$ with the probability
\begin{align*}
\text{P}_{\text{r}}(t,p_{k_*})
 = \frac{\sum_i |w_{ik_*}|^2}{\sum_{i,k} |w_{ik}(t)|^2}
    = \frac{\sum_i |\bm{v}_i(t,p_{k_*})|^2}{\sum_{i,k} |\bm{v}_i(t,p_k)|^2}
    = \frac{ \|\bm{v}(t,p_{k_*})\|^2}{\sum_k \|\bm{v}(t,p_k)\|^2}.
\end{align*}
Let $\{p_k\ge 0\}$ denote the index set of $k$ with $p_k\ge 0$. Then the likelihood of acquiring $\ket{\bm{u}_*}$ that satisfies $p_{k_*} >0$ is given by
\begin{equation} \label{Pr}
\text{P}_{\text{r}}^*
 = \frac{ \sum_{p_k>0} \|\bm{v}(t,p_k)\|^2}{\sum_k \|\bm{v}(t,p_k)\|^2}
 \approx  \frac{\int_0^{\infty} \|\bb{v}(t,p)\|^2 \d p}{\int_{-\infty} ^{\infty} \|\bb{v}(t,p)\|^2 \d p}
\equiv \text{P}_{\text{r}}(t,p\ge 0).
\end{equation}

\begin{lemma} \label{lem:Pr}
Suppose that the real part matrix $H_1$ is negative semi-discrete. Let $\bb{v}(t,p)$ be the solution of \eqref{u2v}. Then the probability on the right-hand side of \eqref{Pr} is
\[\text{P}_{\text{r}}(t,p\ge 0) = \frac{\int_0^{\infty} \|\bb{v}(t,p)\|^2 \d p}{\int_{-\infty} ^{\infty} \|\bb{v}(t,p)\|^2 \d p}
=\frac12 \Big(\frac{\|\bb{u}(t)\|}{\| \bb{u}(0) \|}\Big)^2.\]
This implies a multiplicative factor $g_0 = 2 (\| \bb{u}(0) \|/\|\bb{u}(t)\|)^2$ in the time complexity, which characterizes the decay of the
final state relative to the initial state.
\end{lemma}

\begin{proof}
It is simple to show that the solution of \eqref{u2v} preserves the $L^2$ norm with respect to $p$.
For a given function $v(p)$, we define the normalized Fourier transform of $v(p)$ by $\hat{v}(k)$.  Applying the forward transform to \eqref{u2v}, we get
\[
 \frac{\d}{\d t} \hat{\bb{v}}(t,k) = A \hat{\bb{v}}(t) = -\i k H_1 \hat{\bb{v}} + \i H_2 \hat{\bb{v}}.
\]
 The unitary dynamics implies $\| \hat{\bb{v}}(t,k) \| = \|\hat{\bb{v}}(0,k)\|$ for every $k \in \mathbb{R}$. Since the defined Fourier transform is an isometry of $L^2(\mathbb{R})$, one has
 \[\int_{-\infty}^{\infty} \|\bb{v}(t,p)\|^2 \d p
 = \int_{-\infty}^{\infty} \| \hat{\bb{v}}(t,k) \|^2 \d k
 = \int_{-\infty}^{\infty} \| \hat{\bb{v}}(0,k) \|^2 \d k
 = \int_{-\infty}^{\infty} \|\bb{v}(0,p)\|^2 \d p, \quad t\ge 0.\]
For the initial data in \eqref{u2v}, a direct calculation gives
\begin{align}
\int_{-\infty}^{\infty} \|\bb{v}(t,p)\|^2 \d p
=  \int_{-\infty}^{\infty} \e^{-2 |p|}  \d p \cdot \| \bb{u}_0 \|^2 = \| \bb{u}_0 \|^2. \label{conserv0}
\end{align}

When $\lambda_i\le 0$ for $i=1,\cdots,n$, it is clear that $\bb{v}(t,p) = \e^{-p} \bb{u}(t)$ for $p\ge 0$ (see \cite{JLM24SchrInhom}). This yields
\[\int_0^{\infty} \|\bb{v}(t,p)\|^2 \d p = \int_0^{\infty} \e^{-2p} \d p \|\bb{u}(t)\|^2 = \frac{1}{2} \|\bb{u}(t)\|^2,\]
as required.
\end{proof}

\begin{remark}
We can replace $g_0$ with its square root $\sqrt{g_0}$ using amplitude amplification, as demonstrated in \cite{BerryChilds2017ODE}.
\end{remark}

\section{The conservation form} \label{sec:conservation}

Eq.~\eqref{FP0} can be written in conservation form
\begin{equation}\label{FPConservation}
    \partial_t f = \sigma \nabla \cdot (\e^{-V/\sigma} \nabla (\e^{V/\sigma}  f ) ).
\end{equation}

\subsection{The Schr\"odingerized system}

For the spatial discretization, we consider the Fourier spectral method. To this end, we introduce some notations for later use.
For one-dimensional problems we choose a uniform spatial mesh size $\Delta x = (b-a)/M$ for $M=2N$ an even positive integer and the time step $\Delta t$, and we let the grid points and the time step be
\[x_j = a + j \Delta x, ~~ t_n = n \Delta t, \quad j = 0,1,\cdots, M,~~ n = 0,1,\cdots.\]
For $x\in [a,b]$, the 1-D basis functions for the Fourier spectral method are usually chosen as
\[\phi_l(x) = \e ^{\i \mu_l (x-a)} , \quad \mu_l = \frac{2\pi l}{b-a}, \quad  1 = -N,\cdots, N-1.\]
For convenience, we adjust the index as
\[\phi_l(x) = \e ^{\i \mu_l (x-a)} , \quad \mu_l = \frac{2\pi (l-N-1) }{b-a}, \quad 1 \le l \le M = 2N.\]
The approximation in the 1-D space is
\begin{equation}\label{Fexpand}
u(t,x) = \sum\limits_{l=1}^M c_l(t) \phi_l(x), \quad x = x_j, ~~ j = 0,1,\cdots, M-1,
\end{equation}
where, for simplicity, we still use the notation for the exact solution, and note that the periodic boundary conditions are naturally implied in this expansion, which is also written in vector form,
$\bb{u}(t) = \Phi \bb{c}(t)$,
where
\[\bb{u}(t) = (u(t,x_j))_{M\times 1}, \quad \bb{c} = (c_l)_{M\times 1}, \quad
\Phi = (\phi_{jl})_{M\times M} = (\phi_l(x_j))_{M\times M}.\]

The $d$-dimensional grid points are then given by ${x}_{\bb{j}} = (x_{j_1}, \cdots, x_{j_d})$, where $\bb{j} = (j_1,\cdots,j_d)$, and
\[x_{j_i} = a + j_i \Delta x, \quad j_i = 0,1,\cdots, M-1, \quad i = 1,\cdots,d.\]
We use the notation $ 1\le \bb{j} \le M$ to indicate $1\le j_i \le M$ for every component of $\bb{j}$.
The multi-dimensional basis functions are written as
$\phi_{\bb{l}}({x}) = \phi_{l_1}(x_1)\cdots \phi_{l_d}(x_d)$,
where $\bb{l} = (l_1,\cdots, l_d)$ and $1 \le \bb{l} \le M$. The corresponding approximate solution is
$u(t,{x}) = \sum\nolimits_{\bb{l}} c_{\bb{l}}(t) \phi_{\bb{l}}({x})$,
with the coefficients determined by the exact values at the grid or collocation points ${x}_{\bb{j}} $. These collocation values will be arranged as a column vector:
\[\bb{u}(t) = \sum\limits_{\bb{j}}u(t,{x}_{\bb{j}}) \ket{j_1} \otimes \cdots \otimes \ket{j_d}.\]
That is, the $n_{\bb{j}}$-th entry of $\bb{u}$ is $u(t,{x}_{\bb{j}})$, with the global index given by
\[n_{\bb{j}}: = j_12^{d-1} + \cdots + j_d2^0, \qquad \bb{j} = (j_1,\cdots,j_d). \]
Similarly $c_{\bb{l}}$ is written in a column vector as $\bb{c} = \sum\nolimits_{\bb{l}} c_{\bb{l}} \ket{l_1} \otimes \cdots \otimes \ket{l_d}$. One can get $\bb{u} = \Phi^{\otimes ^d } \bb{c}$.

For later use, we introduce the position operator $\hat{x}_j$ and the momentum operator $\hat{p}_j = -\i \frac{\partial}{\partial x_j}$ in discrete settings.

We first consider the one-dimensional case.
Let $u(x)$ be a function in one dimension and $\bb{u} = [u(x_0),\cdots,u(x_{M-1})]^T$ be the mesh function with $M=2N$. The discrete position operator $\hat{x}^{\rm d}$ of $\hat{x}$ can be defined as
\[\hat{x}^{\rm d}: \bb{u} = \Big( u(x_i) \Big) \quad \to \quad
  \Big( x_i u(x_i) \Big) = D_x \bb{u}  \qquad \mbox{or} \qquad
  \hat{x}^{\rm d}\bb{u} = D_x \bb{u},\]
where $D_x = \text{diag} ( x_0, x_1, \cdots, x_{M-1} )$ is the matrix representation of the position operator in $x$-space. By the discrete Fourier expansion in \eqref{Fexpand}, the momentum operator can be discretized as
\begin{align*}
\hat{p}u(x)
& \approx \hat{p} \sum\limits_{l=1}^M c_l \phi_l(x) = \sum\limits_{l=1}^M c_l \hat{p} \phi_l(x)
  = \sum\limits_{l=1}^M c_l (-\i \partial_x \phi_l(x)) \\
& = \sum\limits_{l=1}^M c_l \mu_l \phi_l(x),  \quad \mu_l = \frac{2\pi (l-N-1) }{b-a}
\end{align*}
for $x = x_j$, $j = 0,1,\cdots, M-1$, which is written in matrix form as
\begin{equation}\label{Pmu}
\hat{p}^{\rm d} \bb{u} =  \Phi D_\mu \Phi^{-1} \bb{u} =: P_\mu \bb{u}, \qquad
D_\mu = \text{diag} ( \mu_1, \cdots, \mu_M ),
\end{equation}
where $\hat{p}^{\rm d}$ is the discrete momentum operator. The matrices $D_\mu$ and $P_\mu$ can be referred to as the matrix representation of the momentum operator in $p$-space and $x$-space, respectively, and are related by the discrete Fourier transform.

For $d$ dimensions, we still denote
$\bb{u} = \sum\nolimits_{\bb{j}} u(x_{\bb{j}}) \ket{j_1} \cdots \ket{j_d}$.
The discrete position operator $\hat{x}_l^{\rm d}$ is defined as
\[\hat{x}_l^{\rm d} \bb{u} = (I^{\otimes^{l-1}} \otimes D_x \otimes I^{\otimes^{d-l}}) \bb{u} =: \bb{D}_l \bb{u}. \]
One easily finds that
\[\hat{P}_l^{\rm d} \bb{u} = (I^{\otimes^{l-1}} \otimes P_\mu \otimes I^{\otimes^{d-l}}) \bb{u} =: \bb{P}_l \bb{u}. \]
Note that
\begin{equation}\label{PlDl}
(\Phi^{\otimes^d})^{-1} \bb{P}_l \Phi^{\otimes^d} = I^{\otimes^{l-1}} \otimes D_\mu \otimes I^{\otimes^{d-l}}=:\bb{D}^\mu _l.
\end{equation}

For the Schr\"odingerization method, we use the warped  transformation $F(t,x,p) = \e^{-p} f(t,x)$ for $p\ge 0$ and extend the initial data to $p<0$ to obtain
\[\begin{cases}
\partial_t F = \sigma \nabla_x \cdot \Big(\e^{-V/\sigma} \nabla_x \Big(\e^{V/\sigma}  (- \partial_p F) \Big) \Big), \\
F(0,x,p) = \e^{-|p|} f(t,x).
\end{cases}\]
Apply the discrete Fourier transformation on $x$ to get
\begin{align*}
& \nabla_x \cdot \Big(\e^{-V/\sigma} \nabla_x \Big(\e^{V/\sigma}  (- \partial_p F) \Big) \Big)\\
&  = \sum\limits_{l=1}^d \partial_{x_l} \Big(\e^{-V/\sigma} \partial_{x_l} \Big(\e^{V/\sigma}  (- \partial_p F) \Big) \Big) \\
& = \sum\limits_{l=1}^d (-\i \partial_{x_l}) \Big(\e^{-V/\sigma} (-\i \partial_{x_l}) \Big(\e^{V/\sigma}  \partial_p F \Big) \Big) \\
& \longrightarrow  \sum\limits_{l=1}^d  \bb{P}_l  \bb{e}^{-\bb{V}/\sigma}  \bb{P}_l\bb{e}^{\bb{V}/\sigma}  \partial_p \bb{F} ,
\end{align*}
where
\[\bb{F} = \bb{F}(t,p) = \sum\limits_{\bb{j}} F(t,x_{\bb{j}},p) \ket{\bb{j}}, \qquad \partial_p \bb{F} = \partial_p \bb{F}(t,p) = \sum\limits_{\bb{j}} \partial_p F(t,x_{\bb{j}},p) \ket{\bb{j}},\]
and $\bb{e}^{\bb{V}} = \text{diag}(\bb{g})$ is a diagonal matrix with the diagonal vector given by
$\bb{g} = \sum_{\bb{j}} \e^{V(x_{\bb{j}})} \ket{\bb{j}}$.
It should be pointed out that $\bb{e}^{\bb{V}}$ is not the matrix exponential $\e^{\bb{V}}$, where we have used the bold symbol $\bb{e}$ to indicate the difference.
Let $\bb{A}_l = \bb{P}_l \bb{e}^{-\bb{V}/\sigma} \bb{P}_l$, which are Hermitian matrices. One has the following PDEs
\begin{equation}\label{FPeq}
\partial_t\bb{F} =  \sigma \sum\limits_{l=1}^d  (\bb{A}_l \bb{e}^{\bb{V}/\sigma}) \partial_p \bb{F} =: - \bb{A}\partial_p \bb{F},
\end{equation}
where
\[\bb{A} = - \sigma \sum\limits_{l=1}^d  \bb{A}_l \bb{e}^{\bb{V}/\sigma}.\]
We can decompose $\bb{A}$ into a Hermitian term and an anti-Hermitian term:
\[\bb{A} = \bb{H}_1 + \i \bb{H}_2,\]
where
\[
\bb{H}_1 = \frac{\bb{A}+\bb{A}^{\dagger}}{2} = \bb{H}_1^{\dagger}, \qquad \bb{H}_2 = \frac{\bb{A}-\bb{A}^{\dagger}}{2 \i} = \bb{H}_2^{\dagger}.
\]
The system is then written as
\begin{equation} \label{quasiHam}
\begin{cases}
\partial_t \bb{F}(t,p) = - \bb{H}_1\partial_p \bb{F}  - \i \bb{H}_2 \partial_p \bb{F}, \\
\bb{F}(0,p) = \e^{-|p|} \bb{F}_0 = \e^{-|p|} \bb{F}(0,p).
 \end{cases}
\end{equation}
Due to the presence of the imaginary unit $\i=\sqrt{-1}$ in the second term, directly applying the discrete Fourier transform to $p$ does not yield a Hamiltonian system. Noting that for $p>0$ there holds
\begin{equation}\label{warpedProperty}
\partial_p \bb{F} = - \bb{F}, \qquad p>0,
\end{equation}
we can modify \eqref{quasiHam} as
\begin{equation} \label{quasiHam-1}
\begin{cases}
\partial_t \bb{F}(t,p) = - \bb{H}_1\partial_p \bb{F}  +  \i \bb{H}_2 \bb{F}, \\
\bb{F}(0,p) = \e^{-|p|} \bb{F}_0 = \e^{-|p|} \bb{F}(0,p),
 \end{cases}
\end{equation}
which is exactly the Schr\"odingerized system for general linear ODEs (see Section \ref{sec:Schr}).

\begin{remark}
In \cite{JLY22SchrLong}, the heat equation and linear convection equations are treated using the same approach as previously described. This involves initially applying the warped phase transformation, followed by spatial discretization. However, when dealing with the Fokker-Planck equation, directly applying the discrete Fourier transform on $p$ does not result in a Hamiltonian system. Nonetheless, we can readily adapt the aforementioned method to establish a Schr\"odingerized system for general ODEs (see Section \ref{sec:Schr}), as indicated by the straightforward observation \eqref{warpedProperty}. Consequently, in various applications, it is preferable to first employ spatial discretization before implementing the Schr\"odingerization method, particularly when non-periodic boundary conditions are imposed. See the next remark.
\end{remark}

\begin{remark}
For non-periodic boundary conditions, let's examine a scenario where $V(x)$ is a constant and $\sigma =1$. In this instance, the Fokker-Planck equation reduces to the heat equation $u_t = u_{xx}$ in one dimension. Let's impose non-homogeneous Dirichlet boundary conditions. If we first apply the warped transformation $v(t,x,p) = \e^{-p} u(t,x)$ to get $\partial_t v = \partial_{xx} (-\partial_p v)$, then discretizing on $x$ using finite differences yields
\[\partial_t \bb{v}(t,p) =  - L_h \partial_p \bb{v} +  \bb{b}(t,p),\]
where $L_h$ represents the discrete Laplacian, and
\[\bb{b}(t,p) = \e^{-p} \tilde{\bb{b}}(t), \qquad \tilde{\bb{b}}(t) = [u(t,-1), 0,\cdots, 0, u(t,1)]^T. \]
Applying the discrete Fourier transform on $p$ gives
\[\frac{\d }{\d t} \bb{w}(t) = \i (\bb{L}_h \otimes P_\mu) \bb{w}(t) + B(t),\]
where $\bb{w}$ is defined as in \eqref{heatww} and $B(t) = \sum_{i,k} \bb{b}_i(t,p_k)\ket{i,k}$, with $M_p$ being the number of grid points for $p$. In this case, one can use  Duhamel's principle to express the solution as
\[\bb{w}(t) = \e^{\i (\bb{L}_h \otimes P_\mu) t } \bb{w}(0) + \int_0^t \e^{\i (\bb{L}_h \otimes P_\mu) s } B(s) \d s \]
and utilize the linear combination of unitaries (LCU) to coherently prepare the state $\bb{w}(t)$.  An alternative approach is to employ the augmentation technique in \cite{JLY22SchrLong} to transform the equation to the case where $B(t) = \bb{0}$. In this case, the evolution matrix will be time-dependent, so one can then apply the quantum simulation technique in \cite{CJL23TimeSchr} for non-autonomous systems, where a non-autonomous system is transferred to an autonomous one in one higher dimension.
\end{remark}

\subsection{Real part of matrix with positive eigenvalues}

 One of the issues to be taken care of  is that $H_1$ can have positive eigenvalues, triggering potential numerical instability. For example, when taking $V(x) = x^2/2$ in one dimension, this can be easily confirmed numerically by checking that the determinant of the second-order leading principal minor is greater than zero:
\[\det\left( \begin{bmatrix}(\bb{H}_1)_{11} &  (\bb{H}_1)_{12} \\ (\bb{H}_1)_{21} & (\bb{H}_1)_{22}\end{bmatrix} \right) > 0.\]

\begin{figure}[H]
  \centering
  \includegraphics[scale=0.5]{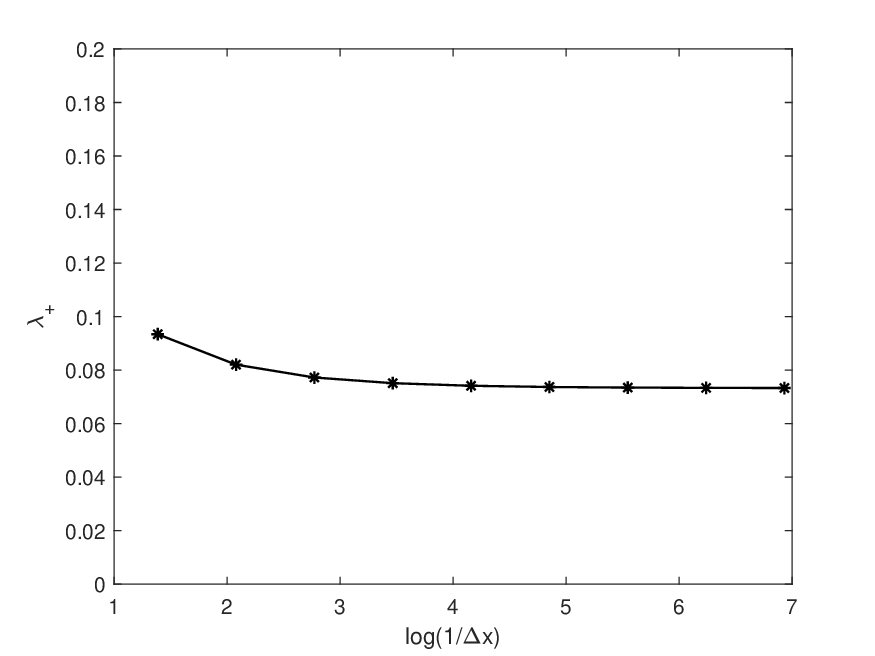}\\
  \caption{The maximum positive eigenvalues vary with different spatial mesh sizes.}\label{fig:lam}
\end{figure}

In Fig.~\ref{fig:lam}, we display the maximum positive eigenvalues for different spatial mesh sizes, where a stable trend ($\lambda_+ \approx 0.073$) can be observed as the mesh size decreases.

This issue was addressed in \cite{JLM24SchrInhom}. The idea there is that to  recover the original variable one has to use $p$ in an appropriate domain.
Without loss of generality, we let $H_2 = O$ and consider
\begin{equation}\label{posu}
\frac{\d \bb{u}(t)}{\d t} =  H_1 \bb{u}(t), \qquad t\ge 0.
\end{equation}
The exact solution is $\bb{u}(t) = \e^{H_1 t} \bb{u}(0)$, or
\[\tilde{\bb{u}}(t) =  \e^{\Lambda t} \tilde{\bb{u}}(0),
\qquad \mbox{i.e.,} \quad  \tilde{\bb{u}}_i(t) =  \e^{\lambda_i t} \tilde{\bb{u}}_i(0), \quad i=1,\cdots,n,\]
where $\tilde{\bb{u}}(t) = Q^{-1}\bb{u}(t)$ and $Q^\dag H_1 Q = \Lambda = \text{diag}(\lambda_1,\cdots,\lambda_n)$ with $Q$ being a unitary matrix.
Using the warped phase transformation $\bb{v}(t,p) = \e^{-p} \bb{u}(t)$ for $p\ge 0$ and symmetrically extending the initial data to $p<0$, one has
\begin{equation}\label{posv}
\begin{cases}
\frac{\d }{\d t}  \bb{v}(t,p)=  - H_1 \partial_p \bb{v}(t,p), \quad t\ge 0,\\
\bb{v}(0,p) = \e^{-|p|} \bb{u}(0).
\end{cases}
\end{equation}
The diagonalized system is
\[\begin{cases}
\frac{\d }{\d t} \tilde{\bb{v}}(t,p) =  - \Lambda \partial_p \tilde{\bb{v}}(t,p), \\
\tilde{\bb{v}}(0,p) =  \e^{-|p|}\tilde{\bb{u}}(0),
\end{cases}\]
where $\tilde{\bb{v}}(t,p) = Q^{-1}\bb{v}(t,p)$. Evidently, the solution is
\begin{equation}\label{vtilsol}
\tilde{\bb{v}}_i(t,p) = \tilde{\bb{v}}_i(0,p-\lambda_i t) = \e^{-|p-\lambda_i t|}\tilde{\bb{u}}_i(0), \qquad i=1,\cdots,n.
\end{equation}
By using this formula, it is simple to prove that we can recover the solution in the range of $p \ge \max \{ \lambda_+ t ,0 \}$, where $\lambda_+$ denotes the maximum value among the positive eigenvalues of $H_1$. For general case where $H_2 \ne 0$ ($H_1$ and $H_2$ can be time dependent), one can refer to \cite{JLM24SchrInhom}, which uses the recovery strategy as in the next theorem.

\begin{theorem} \label{thm:gplus}
Suppose that $\bb{u}(t)$ and $\bb{v}(t)$ are the solutions of \eqref{posu} and \eqref{posv}, respectively.  Let $\lambda_+$ denote the maximum value among the positive eigenvalues of $H_1$. Then the solution can be restored as
\begin{equation}\label{restore}
\bb{u}(t) = \e^{p}\bb{v}(t,p), \qquad p \ge \max \{ \lambda_+ t ,0 \}.
\end{equation}
\end{theorem}

With this, one can get the corresponding state vector in this range with the probability given by
\[
\text{P}_{\text{r}}(t,p\ge \lambda_+ t ) = \frac{\int_{\lambda_+ t}^{\infty} \|\bb{v}(t,p)\|^2 \d p}{\int_{-\infty} ^{\infty} \|\bb{v}(t,p)\|^2 \d p} = \frac{1}{2} \Big(\frac{\|\e^{- \lambda_+ t} \bb{u}(t)\|}{\| \bb{u}(0) \|}\Big)^2.
\]
This implies a multiplicative factor
 \[g_+  =  2  \Big(\frac{\|\bb{u}(0)\|}{\| \e^{-  \lambda_+ t} \bb{u}(t) \|}\Big)^2\]
in the time complexity.

If $H_1$ has positive eigenvalues, one can also define $\bb{z}(t) = \e^{-ct} \bb{u}(t)$ as in \cite{ALL2023LCH} and rewrite \eqref{ODElinear} as
\begin{equation}\label{echange}
 \begin{cases}
 \frac{\d \bb{z}(t)}{\d t} = (H_1 - c I) \bb{z}(t) + \i H_2 \bb{z}(t), \\
 \bb{z}(0) = \bb{u}_0.
 \end{cases}
\end{equation}
It is evident that the matrix $H_1 - cI$ for the non-unitary part is negative semi-definite if $c = \lambda_{\max}(H_1)$.  For the transformed equation \eqref{echange}, the application of the Schr\"odingerization technique ensures the restoration of the correct solution for all $p\ge 0$.  According to Lemma \ref{lem:Pr}, the multiplicative factor in the time complexity for \eqref{echange}  is adjusted to
\begin{equation}\label{gc}
g_c = 2  \Big(\frac{\|\bb{z}(0)\|}{\| \bb{z}(t) \|}\Big)^2 = 2  \Big(\frac{\|\bb{u}(0)\|}{\| \e^{-ct}\bb{u}(t) \|}\Big)^2,
\end{equation}
where $c = \lambda_{\max}(H_1) = \lambda_+$ with $\lambda_+$ denoting the maximum value among the positive eigenvalues of $H_1$.
It can be observed that $g_c = g_+$. This indicates that the exponential change of variables ($\bb{z}(t) = \e^{-ct} \bb{u}(t)$) has the same multiplicative factor in time complexity as the  Schr\"odingerization approach when projecting onto the solution state.

\begin{remark} \label{rem:equiv1}
 The conclusion in Theorem \ref{thm:gplus} is for the continuous Schr\"odingerization (see \cite{JLM24SchrInhom}).
As depicted in Section \ref{sec:Schr} (see also Fig.~\ref{fig:domain}), when discretizing in the $p$ direction by the discrete Fourier transformation, it is crucial to select the left endpoint $L$ in such a way that the fastest left-moving wave does not reach $p=L<0$ during the duration of the computation. This requirement can be expressed as
\[
s_* T \le L_0 - L \qquad \mbox{or} \qquad L \ge L_0 - s_* T,
\]
where $s_* = |\lambda_{\min}|$ is the speed for the fastest left-moving wave (Note that $\lambda_{\min} < 0$). Similarly, when the matrix $H_1$ has positive eigenvalues, the right endpoint $R$ should satisfy $s^* T \le R$ and the periodic boundary condition $\e^R \approx 0$, where $s^* = \lambda_{\max} = \lambda_+>0$ represents the speed of the fastest right-moving wave.
\end{remark}

\subsection{A symmetric transformation}  \label{subsubsec:symmetric}

Instead of the decomposition of $\bb{A}$ in \eqref{FPeq}, we can define $\tilde{\bb{F}}(t,p) = \bb{e}^{{\bb{V}}/(2\sigma)} \bb{F}(t,p)$ and get a symmetric form
\begin{equation}\label{quasiHamSym}
\partial_t \tilde{\bb{F}} =   - \bb{H}  \partial_p \tilde{\bb{F}},
\end{equation}
where
\[\bb{H}  =    - \sigma \bb{e}^{{\bb{V}}/(2\sigma)} ( \bb{A}_1 + \cdots + \bb{A}_d) \bb{e}^{{\bb{V}}/(2\sigma)} ,  \qquad \bb{A}_l = \bb{P}_l \bb{e}^{-\bb{V}/\sigma} \bb{P}_l.\]
Eq.~\eqref{quasiHamSym} gives a well-defined Schr\"odingerized system
\begin{equation}\label{HamSym}
\frac{\d}{\d t} \bb{w}(t) = -\i ( \bb{H} \otimes P_\mu ) \bb{w} \qquad \mbox{or} \qquad
\frac{\d}{\d t} \tilde{\bb{w}}(t) = -\i ( \bb{H} \otimes D_\mu ) \tilde{\bb{w}}
\end{equation}
for all $p\ge 0$ since $\bb{H}$ is negative semi-definite, where $\bb{w} = \sum_{\bb{j}, k} \bb{F}(t,x_{\bb{j}}, p_k) \ket{\bb{j}}\ket{k}$ and $\tilde{\bm{w}} = (I \otimes \Phi^{-1})\bm{w}$. In fact, for any vector $\bb{x} \ne \bb{0}$ and $\bb{z} = \bb{e}^{{\bb{V}}/(2\sigma)} \bb{x}$, noting that $\bb{P}_l^\dag = \bb{P}_l$, one has
\begin{align*}
\bb{x}^\dag \bb{H}\bb{x}
 =    - \sigma \sum\limits_{l=1}^d  \bb{z}^\dag  \bb{A}_l \bb{z}
  = - \sigma  \sum\limits_{l=1}^d \bb{z}^\dag \bb{P}_l \bb{e}^{-\bb{V}/\sigma} \bb{P}_l \bb{z} \equiv
  - \sigma \sum\limits_{l=1}^d \bb{y}_l^\dag \bb{e}^{-\bb{V}/\sigma} \bb{y}_l \le 0.
\end{align*}
Although the matrix $\bb{H}$ can be efficiently prepared on a classical computer by using the fast Fourier transform (FFT), it should be pointed out that  $\bb{H}$ is not sparse  along the $x_l$ direction. This implies an additional multiplicative factor of $ d M_x $ in the complexity, where $M_x$ is the number of grid points in one dimension.

Given the absence of any special structure in the evolution matrix that would facilitate diagonalization, we shall utilize general methods for Hamiltonian simulations.
The quantum algorithm for general sparse Hamiltonian simulation with nearly optimal dependence on all parameters can be found in \cite{Berry-Childs-Kothari-2015}.
\begin{lemma}
A time-independent Hamiltonian $H$ acting on $m$ qubits can be simulated within error $\varepsilon$ with time complexity
\begin{equation}\label{time}
\mathcal{O}( \tau   m \cdot L_{\text{polylog}}), \qquad L_{\text{polylog}}  \equiv  (1+\log^{2.5}(\tau/\varepsilon) )\frac{\log (\tau/\varepsilon) }{\log\log (\tau/\varepsilon)},
\end{equation}
where $\tau = s \|H\|_{\max} T$, $s$ is the sparsity of $H$ and $T$ is the evolution time.
\end{lemma}

\begin{figure}[H]
  \centering
  \includegraphics[scale=0.2]{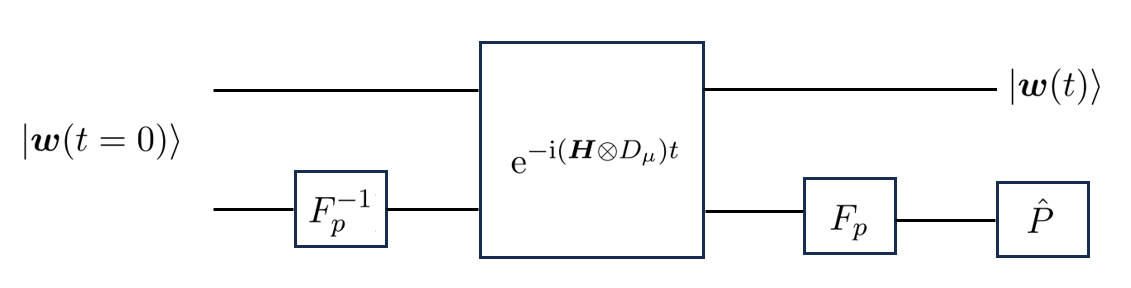}\\
  \caption{Quantum circuit for the conservation form. The black box for the simulation of $\bb{H}\otimes D_\mu$ can be chosen as the method given in \cite{CJL23TimeSchr}.}\label{fig:Conservationsym}
\end{figure}

For time-dependent Hamiltonian $H(t)$, one can use the result by Berry et al. \cite[Theorem 10]{BerryChilds2020TimeHamiltonian}, where complicated procedures involving the Dyson series are required, with the time complexity given by
$\widetilde{\mathcal{O}}( s \norm{H}_{\max,1} m)$, where $\widetilde{\mathcal{O}}$ suppresses all logarithmic factors and the norm is defined as
\begin{equation*}
    \norm{H}_{\max,1} =\int_0^T \norm{H}_{\max}(t) \d t, \quad
\norm{H}_{\max}(t):= {\max}_{i,j} \abs{H_{i,j}(t)}.
\end{equation*}
One can also use the recent results in \cite{CJL23TimeSchr}, in which the authors proposed an alternative formalism that turns any non-autonomous unitary dynamical system into an autonomous unitary system, i.e., quantum system with a time-independent
Hamiltonian, in one higher dimension. The time complexity is dominated by $\mathcal{O}(s  \max_{t \in [0,T]} \|H(t)\|_{\max} T )$ (see Remark 16 there).  The quantum circuit for solving \eqref{quasiHam-1} or \eqref{HamSym} is displayed in Fig.~\ref{fig:Conservationsym}: Given $\ket{\bb{w}(0)}$, we perform the inverse quantum Fourier transform concerning $p$ to obtain $\ket{\tilde{\bb{w}}(0)}$. After applying
the quantum simulation method in \cite{CJL23TimeSchr} for example, we acquire $\ket{\tilde{\bb{w}}(t)}$. To retrieve $\ket{\bb{w}(t)}$, we employ the quantum Fourier transform on the $p$ mode. As discussed in the previous section, only the parts of the auxiliary mode where $p\ge \lambda_+ t$ are necessary to recover $\ket{\bb{\psi}(t)}$, where $\lambda_+>0$ for \eqref{quasiHam-1} and $\lambda_+ = 0$ for \eqref{HamSym}. Thus, in a basic setup, we can use $\hat{P}$ to project out solely the $p\ge \lambda_+ t$ components from the auxiliary mode.

Based on the preceding discussions, we have identified two approaches for solving the conservation form \eqref{FPConservation}, which we categorize as follows:
\begin{itemize}
  \item Conservation I: Utilizing the Schr\"odingerization directly for \eqref{quasiHam-1}.
  \item Conservation II: Applying the symmetric transformation in \eqref{quasiHamSym}.
\end{itemize}
The complexities associated with these methods are summarized in the following theorem.

\begin{theorem} \label{thm:conservationTime}
Suppose $|V(x)| \le C$ is bounded, and the conditions in Remark \ref{rem:equiv1} for the left and right endpoints $L$ and $R$ are satisfied. Let $g_0 = (\|\bb{f}(0)\|/\|\bb{f}(T)\|)^2$. Consider $\bb{f}(t)$ as the exact solution to the semi-discrete problem of \eqref{FPConservation}, and let $\tilde{\bb{f}}(t) = \sum_{\bb{j}}\e^{V(x_{\bb{j}})/(2\sigma)} f(t,x_{\bb{j}}) \ket{\bb{j}}$.
\begin{enumerate}[1)]
  \item For Conservation II, the Schr\"odingerization approach can produce a state $\varepsilon$-close to $\bb{f}(T)/\|\bb{f}(T)\|$ in $l^2$ norm, with time complexity
\[\mathcal{Q}_{\text{Conservation-I}} = \e^{2 \lambda_+ T} g_0 \cdot \widetilde{\mathcal{O}}\left( \frac{d^{3+5/\ell} T^2}{\varepsilon^{1+5/\ell}} + \frac{d^{2+3/\ell} R T}{\varepsilon^{1+3/\ell}} \right),\]
 where $\lambda_+$ is the maximum eigenvalue of $\bb{H}_1$ in \eqref{quasiHam}.
  \item For Conservation II, the Schr\"odingerization approach can produce a state $\varepsilon$-close to $\tilde{\bb{f}}(T)/\|\tilde{\bb{f}}(T)\|$ in $l^2$ norm, with time complexity
\[\mathcal{Q}_{\text{Conservation-II}} = g_0 \cdot \widetilde{\mathcal{O}}\left( \frac{d^{3+5/\ell} T^2}{\varepsilon^{1+5/\ell}} + \frac{d^{2+3/\ell} R T}{\varepsilon^{1+3/\ell}}\right).\]
\end{enumerate}
Here, $\ell$ is the Sobolev regularity of $f(t,x)$ (namely $f \in H^\ell$ for fixed $t$).
\end{theorem}
\begin{proof} We only provide the details for the second method.
Since $\bb{H}$ is not sparse along $x_l$ direction, the sparsity $s = \mathcal{O}(d M_x)$.
With direct calculations, one finds that
\begin{align*}
\|\bb{H} \otimes P_\mu \|_{\max}
& = \| - \sigma \bb{e}^{{\bb{V}}/(2\sigma)} ( \bb{A}_1 + \cdots + \bb{A}_d) \bb{e}^{{\bb{V}}/(2\sigma)} \|_{\max}\cdot \|P_\mu \|_{\max}  \\
& \le \sigma \|\bb{e}^{{\bb{V}}/(2\sigma)}\|_{\max}^2  \|\bb{e}^{-\bb{V}/\sigma}\|_{\max} (\|\bb{P}_1\|_{\max}^2 + \cdots + \|\bb{P}_d\|_{\max}^2) \|P_\mu \|_{\max}
 \lesssim  d M_x^2 M_p,
\end{align*}
where $M_x$ and $M_p$ are the numbers of grid points along $x$ and $p$ directions.
The error scales as $\mathcal{O}(d \Delta x^\ell  + \Delta p)$, where $\ell$ stands for the regularity in the $x$ direction. The limitation of the convergence order mainly comes from the non-smoothness of the initial value in the $p$ direction.
So we can choose $d \Delta x ^\ell \sim \varepsilon$ and $\Delta p \sim \varepsilon$.

For $d$ dimensions, the fastest left moving wave will have a speed $s_*= \mathcal{O}(d M_x^2)$, so the range of the auxiliary variable $p$ is
\[R-L = R-L_0 + s_* T \sim R-L_0 + d M_x^2 T \sim  R-L_0 + d M_x^2 T \sim R-L_0 + d (d / \varepsilon)^{2 /\ell}T ,\]
where $L_0$ can be chosen as a constant and $\Delta x \sim (\varepsilon/d)^{\ell}$ is used. This gives
\[M_p = (R-L)/\Delta p \sim R/\varepsilon + (d /\varepsilon)^{1+2/\ell} T.\]
By using Lemma \ref{lem:Pr} and plugging the bounds of $s$ and  $\bb{H} \otimes P_\mu $ in \eqref{time}, we obtain
\[\mathcal{Q}_{\text{Conseration-II}} = \tilde{g}_0 \cdot \widetilde{\mathcal{O}}\left( \frac{d^{3+5/\ell} T^2}{\varepsilon^{1+5/\ell}} + \frac{d^{2+3/\ell} R T}{\varepsilon^{1+3/\ell}}\right) ,\]
where
\[\tilde{g}_0 = \Big(\frac{\|\tilde{\bb{f}}(0)\|}{\|\tilde{\bb{f}}(T)\|}\Big)^2 \le \e^{2(V_{\max}-V_{\min})} \Big(\frac{\|\bb{f}(0)\|}{\|\bb{f}(T)\|} \Big)^2 \le C g_0,  \]
as required.
\end{proof}

\begin{remark}
One can use smoother initial data as in \cite{JLM24SchrBackward} to get second-order accuracy in $p$. In this case, selecting $\Delta p$ on the order of $\varepsilon^{1/2}$ results in a multiplicative factor $\varepsilon^{1/2}$  in the complexities mentioned above.
\end{remark}

\begin{remark}
For the second method (Conservation II), after performing measurements in the computational basis of the $p$-register, the solution state vector can be reconstructed as
\[\ket{\tilde{\bb{f}}(t)} = \frac{1}{\mathcal{N}}
\sum_{\bb{j}}\e^{V(x_{\bb{j}})/(2\sigma)} f(t,x_{\bb{j}}) \ket{\bb{j}}, \qquad \mathcal{N} = \sum_{\bb{j}} | \e^{V(x_{\bb{j}})/(2\sigma)} f(t,x_{\bb{j}}) |^2. \]
With this reconstructed vector, various physical quantities can be computed.
For instance, the integral $\int_{\Omega} f(t,x)^2 \d x$ can be approximated by using the numerical quadrature rule:
\begin{align*}
\int_{\Omega} f(t,x)^2 \d x
& = \int_{\Omega} \e^{-V/\sigma} \Big(\e^{V/(2\sigma)}  f \Big)^2  \d x =: \int_{\Omega} \e^{-V/\sigma} \tilde{f}^2  \d x \\
& \approx \sum\limits_{\bb{j}} w_{\bb{j} } \e^{-V(x_{\bb{j}})/\sigma} |\tilde{f}(t,x_{\bb{j}})|^2
=: \sum\limits_{\bb{j}} \tilde{w}_{\bb{j} } |\tilde{f}(t,x_{\bb{j}})|^2,
\end{align*}
where $\tilde{w}_{\bb{j}} = \e^{-V(x_{\bb{j}})/\sigma} w_{\bb{j} }$ denote the scaled weights. The summation will be further expressed as the expectation value of an observable. For details, readers are referred to \cite{JinLiu2022nonlinear,JLY22nonlinear}.
\end{remark}

\section{The  heat equation form} \label{sec:FKheat}

In Section \ref{subsubsec:symmetric}, we utilize a transformation or change of variables to convert the semi-discrete system \eqref{FPeq} into a symmetric form. Hereafter, we implement this transformation directly  to the original problem, resulting in the heat equation form. In fact, let $\psi(t,x)=\e^{V/(2\sigma)}f $, one easily gets that $\psi$ satisfies \cite{MarKVill}
\begin{equation}\label{FPS}
    \partial_t \psi = \sigma \Delta \psi - U(x)\psi,
\end{equation}
where
\[U(x):= \frac{|\nabla V|^2}{4\sigma}-\frac{1}{2} \Delta V.\]

\subsection{Quantum simulation via Schr\"odingerization}

Similarly, we introduce the warped phase transformation $\Psi(t,x,p) = \e^{-p} \psi(t,x)$ for $p\ge 0$ and extend the initial data to $p<0$ to get
\begin{equation}\label{FKheat}
\begin{cases}
\frac{\d}{\d t} \widetilde{\bb{\Psi}}(t) = \i \bb{H}  \widetilde{\bb{\Psi}}(t) , \\
\widetilde{\bb{\Psi}}(0) = (I \otimes F_p^{-1}) \bb{\Psi}(0),
\end{cases}
\end{equation}
where
\[\widetilde{\bb{\Psi}} = (I \otimes F_p^{-1}) \bb{\Psi}, \qquad  \bb{\Psi} = \bb{\Psi}(t,p) = \sum\limits_{\bb{j}} \Psi(t,x_{\bb{j}},p) \ket{\bb{j}},\]
and
\[\bb{H} = ( \sigma(\bb{P}_1^2 + \cdots + \bb{P}_d^2) + \bb{U} ) \otimes D_\mu  \]
is a Hermitian matrix with $\bb{U}$ defined as $\bb{V}$. By comparing with Eq.~\eqref{generalSchr}, we find
\[H_1 = - [\sigma(\bb{P}_1^2 + \cdots + \bb{P}_d^2) + \bb{U}], \quad H_2 = O.\]
Clearly, $U(x)\ge 0$ leads to the negative semi-definiteness of $H_1$.
The Schr\"odingerized  system \eqref{FKheat} can be efficiently solved by the quantum Fourier transform (QFT) based on the time-splitting approximations. We follow the approach of \cite{JLM24SchrInhom} to recover $\psi$.

From time $t=t_n$ to time $t = t_{n+1}$, the system can be solved in two steps:
\begin{itemize}
  \item One first solves
  \[\begin{cases}
  \frac{\d}{\d t}  \widetilde{\bb{\Psi}} (t) = \i ( \sigma ( \bb{P}_1^2 + \cdots + \bb{P}_d^2 ) \otimes D_\mu)   \widetilde{\bb{\Psi}} (t), \qquad t_n < t < t_{n+1}, \\
   \widetilde{\bb{\Psi}} (t_n) =  \widetilde{\bb{\Psi}} ^n
  \end{cases}\]
for one time step, where $\widetilde{\bb{\Psi}}^n$ is the numerical solution at $t=t_n$. By letting $\widetilde{\bb{c}}(t) = ( F_x^{-1} \otimes I) \widetilde{\bb{\Psi}} (t)$, we instead solve
\[\begin{cases}
\frac{\d}{\d t} \widetilde{\bb{c}}(t) = \i \bb{H}_D  \widetilde{\bb{c}}(t),\qquad t_n < t < t_{n+1}, \\
\widetilde{\bb{c}}(t_n) = \widetilde{\bb{c}}^n = (F_x^{-1} \otimes I) \bb{\Psi}^n,
\end{cases}\]
  where
  \[\bb{H}_D = \sigma  ( (\bb{D}_1^\mu)^2 + \cdots + (\bb{D}_d^\mu)^2 ) \otimes D_\mu\]
  is a diagonal matrix. The numerical solution will be denoted by $\widetilde{\bb{c}}^*$.
  \item Let $ \widetilde{\bb{\Psi}}^* = ( F_x \otimes I) \widetilde{\bb{c}}^*$. The second step is to solve
  \[\frac{\d}{\d t} \widetilde{\bb{\Psi}} (t) =  \i (\bb{U}  \otimes D_\mu)   \widetilde{\bb{\Psi}} (t) =: \i \bb{H}_U  \widetilde{\bb{\Psi}} (t) \]
 for one time step, with $ \widetilde{\bb{\Psi}} ^*$ being the initial data, where $\bb{H}_U $ is a diagonal matrix. This gives the updated numerical solution $ \widetilde{\bb{\Psi}}^{n+1}$.
\end{itemize}

Given the initial state of $\widetilde{\bb{\Psi}}^0$, applying the inverse QFT to the $x$-register, one gets $\widetilde{\bb{c}}^0$.
At each time step, one needs to consider the following procedure
\[\widetilde{\bb{c}}^n
\xrightarrow {\e^{\i \bb{H}_D \Delta t}} \widetilde{\bb{c}}^*
\xrightarrow {F_x\otimes I } \widetilde{\bb{\Psi}}^*
\xrightarrow {\e^{\i \bb{H}_U \Delta t}} \widetilde{\bb{\Psi}}^{n+1}
 \xrightarrow {F_x^{-1}\otimes I } \widetilde{\bb{c}}^{n+1},\]
where $F_x = \Phi^{\otimes^d}$. Let $U_{\Delta t} = (F_x^{-1}\otimes I)\e^{\i \bb{H}_U \Delta t} (F_x\otimes I) \e^{\i \bb{H}_D \Delta t}$.

\begin{figure}[!htb]
  \centering
  \includegraphics[scale=0.2]{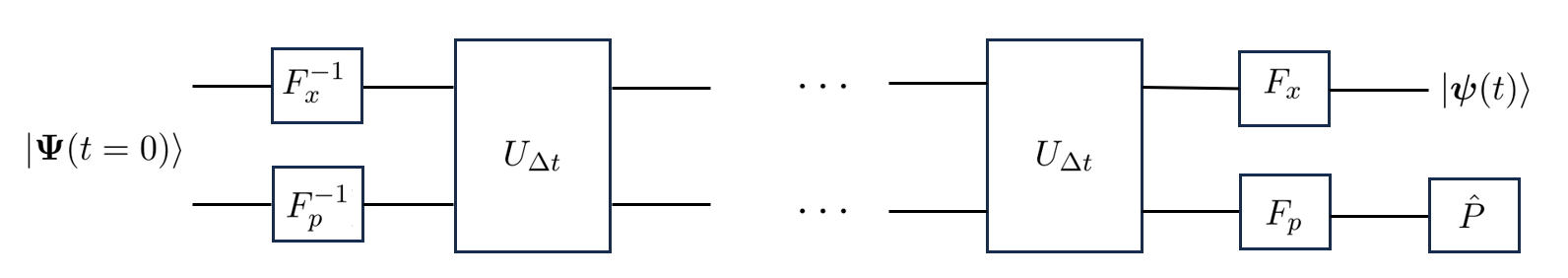} \\
  \includegraphics[scale=0.25]{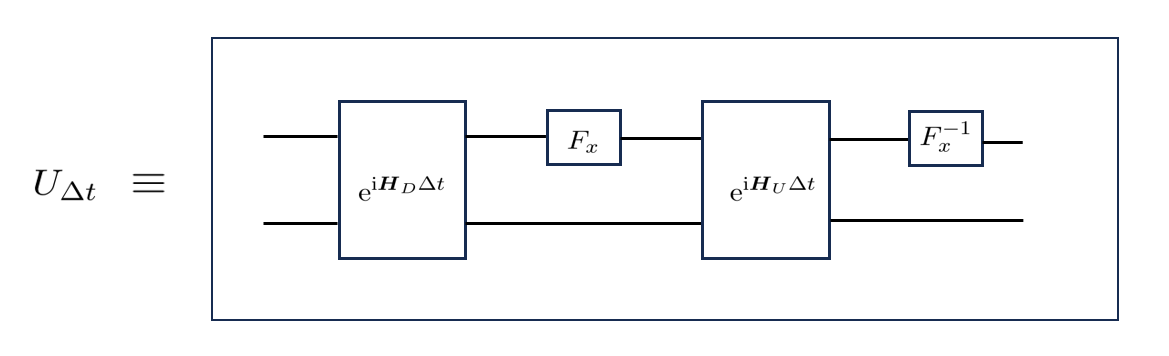}\\
  \caption{Quantum circuit for solving \eqref{FPS}. Top: circuit for the Schr\"odingerization method; Bottom: circuit for the operator $U_{\Delta t}$ at each time step.}\label{fig:circuitheat}
\end{figure}

The quantum circuit depicted in Fig.~\ref{fig:circuitheat} is utilized as follows: Given $\ket{\bb{\psi}(0)}$, we construct a product state $\ket{\bb{\Psi}(t=0)}$ involving the auxiliary mode. Subsequently, we perform the inverse quantum Fourier transform concerning $x$ and $p$ to obtain $\ket{\widetilde{\bb{c}}^0}$. After applying $U_{\Delta t}$ for $N_t$ iterations, we acquire $\ket{\widetilde{\bb{c}}^{N_t}}$. To retrieve $\ket{\bb{\psi}(t)}$, we employ the quantum Fourier transform on the $x$ and $p$ modes, resulting in $\ket{\bb{\Psi}(t)}$. As discussed in the previous section, only the parts of the auxiliary mode where $p\ge \lambda_+ t$ are necessary to recover $\ket{\bb{\psi}(t)}$. Thus, in a basic setup, we can use $\hat{P}$ to project out solely the $p\ge \lambda_+ t$ components from the auxiliary mode.

\begin{theorem}  \label{thm:heatformspectral}
Let $\bb{f}(t)$ be the solution to the semi-discrete problem of \eqref{FPS} and $\tilde{\bb{f}}(t)$ is defined as in Theorem \ref{thm:conservationTime}. Then the Schr\"odingerization approach can produce a state $\varepsilon$-close to $\tilde{\bb{f}}(T)/\|\tilde{\bb{f}}(T)\|$ in $l^2$ norm, with time complexity
\[N_{\text{Gates}} = \e^{2 U_{\max} T} g_0 \cdot \mathcal{O}\left(  \frac{d}{\varepsilon} \log \frac{d^{1/\ell} }{\varepsilon^{1/\ell}} \log\log \frac{d^{1/\ell} }{\varepsilon^{1/\ell}} + \log \Big( \frac{d^{1 + 2/\ell} }{\varepsilon^{1+ 2/\ell}} T + \frac{R}{\varepsilon} \Big) \right),\]
where $g_0$ is defined in Theorem \ref{thm:conservationTime},  $U_{\max} = \max \{ U(x) \}$ and $\ell$ is the Sobolev regularity of $f(t,x)$.
\end{theorem}
\begin{proof}
The error for the time-splitting spectral discretization is $\mathcal{O}(\Delta t + d \Delta x^\ell + \Delta p)$, so we can choose
$\Delta t \sim \varepsilon$, $d \Delta x ^\ell \sim \varepsilon$ and $\Delta p \sim \varepsilon$. As analyzed in the proof of Theorem \ref{thm:conservationTime}, the number of grid in the $p$ direction satisfies
\[M_p = (R-L)/\Delta p \sim R/\varepsilon + (d /\varepsilon)^{1+2/\ell} T\]
for a sufficiently small $\varepsilon$. The numbers of qubits in the $x_i$ ($1\le i \le d)$ and $p$ directions are $n_x \sim \log M_x$ and $n_p \sim \log M_p$.

According to \cite{Kassal2008Diagonal,Jin2022quantumSchrodinger}, the QFT and the diagonal unitary operators with $m$ qubits in each direction can be implemented using $dm$ and $dm \log m$ gates. Thus the gate complexity required to iterate to the $N_t$-th step is
\begin{align*}
N_{\text{Gates}}
 \sim N_t  ( 2d n_x\log n_x +  2 d n_x  ) +  2 ( dn_x + n_p).
\end{align*}
The proof is completed by plugging the bounds of $n_x$ and $n_p$.
\end{proof}

\subsection{The time-splitting Schr\"odingerization method}

The time-splitting algorithm is extensively employed in quantum computing for solving differential equations. This technique effectively harnesses the quantum computational advantages offered by various operators, leading to the development of more practical and comprehensive quantum algorithms. Analogous to quantum simulations in the heat equation form, this approach aims to achieve enhanced computational efficiency. In the following, we focus on the relationship between operator splitting in the Schr\"odingerization method \eqref{u2valpha} and operator splitting in the original problem \eqref{ODElinear}.

For the Schr\"odingerization method \eqref{u2valpha}, we can use the warped phase transformation $\bb{v}(t,p) = \e^{-p} \bb{u}(t)$ for $p\ge 0$ and symmetrically extend the initial data to $p<0$ at the initial moment. Therefore, the time-splitting Schr\"odingerized system can be written as follows.
\begin{itemize}
  \item The first step is to solve
\begin{equation}\label{split1}
\begin{cases}
 \frac{\d}{\d t} \bb{v}^{m,1}(t,p) =  - H_1 \partial_p \bb{v}^{m,1}(t,p), \qquad t \in [t_m, t_{m+1}], \\
 \bb{v}^{m,1}(t_m,p) = \bb{v}^m(p)
 \end{cases}
\end{equation}
from time $t=t_m$ to time $t=t_{m+1}$, where $\bb{v}^m$ is the numerical solution at $t=t_m$ ($m \ge 0$) and  $\bb{v}^0(p) = \e^{-|p|} \bb{u}^0$.
  \item The second step is to solve
  \begin{equation}\label{split2}
\begin{cases}
 \frac{\d}{\d t} \bb{v}^{m,2}(t,p) =  \i H_2 \bb{v}^{m,2}(t,p), \qquad t \in [t_m, t_{m+1}],  \\
 \bb{v}^{m,2}(t_m,p) = \bb{v}^{m,1}(t_{m+1}, p),
 \end{cases}
\end{equation}
again for one time step, with the updated solution denoted by $\bb{v}^{m+1}(p)$.
\end{itemize}

Accordingly, one can consider the resolution of the original problem\eqref{ODElinear} by using the time-splitting method in two steps:
\begin{itemize}
  \item We first solve
  \begin{equation}\label{ODEStep1}
 \begin{cases}
 \frac{\d}{\d t}\bb{u}^{m,1}(t) =  H_1 \bb{u}^{m,1}(t), \qquad t \in [t_m, t_{m+1}], \\
 \bb{u}^{m,1}(t_m) = \bb{u}^m,
 \end{cases}
\end{equation}
from time $t=t_m$ to time $t=t_{m+1}$, where $\bb{u}^m$ is the initial data (at $t = t_m$).

  \item The second step is to solve
    \begin{equation}\label{ODEStep2}
 \begin{cases}
 \frac{\d}{\d t}\bb{u}^{m,2}(t) =  \i H_2 \bb{u}^{m,2}(t), \qquad t \in [t_m, t_{m+1}], \\
 \bb{u}^{m,2}(t_m) = \bb{u}^{m,1}(t_{m+1}),
 \end{cases}
\end{equation}
again for one time step. This gives the updated solution $\bb{u}^{m+1}$.
\end{itemize}

\begin{theorem}\label{thm:validility}
Let $\bb{v}^{m+1}(p)$ be the updated solution of \eqref{split2} in the $m$-th step ($m\ge 0$). If $H_1$ is negative semi-definite, then
\[\bb{v}^{m+1}(p) = \e^{-p} \bb{u}^{m+1}, \qquad p \ge 0,\]
where $\bb{u}^{m+1}$ is the updated solution of \eqref{ODEStep2}.
\end{theorem}
\begin{proof}
We establish this by employing mathematical induction.

Let $Q^\dag H_1 Q = \Lambda = \text{diag}(\lambda_1,\cdots,\lambda_n)$, with $Q$ being a unitary matrix and $\lambda_i \le 0$. Let $\tilde{\bb{v}}^{m,1}(s,p) = Q^\dag\bb{v}^{m,1}(t,p)$, where $t = t_m+s$ and $s\in [0,\Delta t]$.

(1) For the base case $m=0$, Eq.~\eqref{split1} can be expressed as
\[
\begin{cases}
 \frac{\d}{\d s} \tilde{\bb{v}}^{0,1}(s,p) =  - \Lambda \partial_p \tilde{\bb{v}}^{0,1}, \quad s \in [0, \Delta t],\\
 \tilde{\bb{v}}^{0,1}(0,p) = \e^{-|p|} (Q^\dag\bb{u}^0).
 \end{cases}
\]
The solution is straightforwardly given by:
\[\tilde{\bb{v}}_i^{0,1}(s,p) = \e^{- |p-\lambda_i s |} (Q^\dag\bb{u}^0)_i, \qquad i = 1,\cdots,n.\]
Let $\Lambda_{p,s}^0$ denote the diagonal matrix defined as $\{|p-\lambda_i s | :  i=1,\cdots,n\}$. Then,
\[\bb{v}^{0,1}(t,p) = Q \e^{- \Lambda^0_{p,s}} Q^\dag\bb{u}^0.\]
Specifically, for $p\ge 0$, considering that $-|p-\lambda_i s | = -p + \lambda_i s $, we have
\begin{equation}\label{0pg0}
\bb{v}^{0,1}(t,p) = \e^{- p} Q\e^{\Lambda s} Q^\dag\bb{u}^0
= \e^{- p} \e^{H_1 s} \bb{u}^0 = \e^{-p} \bb{u}^{0,1}(t), \quad p \ge 0.
\end{equation}

The solution to \eqref{split2} for the second step is then given by
\[\bb{v}^{0,2}(t,p) = \e^{\i H_2 s}\bb{v}^{0,1}(t_1,p)
= \e^{\i H_2 s} Q \e^{- \Lambda^0_{p,\Delta t}} Q^\dag\bb{u}^0. \]
When $t = t_1$, we have
\[\bb{v}^1(p) = \bb{v}^{0,2}(t_1,p) = \e^{\i H_2 \Delta t} Q \e^{- \Lambda^0_{p,\Delta t}} Q^\dag\bb{u}^0, \]
and for $p\ge 0$, using \eqref{0pg0}:
\[\bb{v}^1(p) = \bb{v}^{0,2}(t_1,p) = \e^{\i H_2 \Delta t}\bb{v}^{0,1}(t_1,p)
 =  \e^{-p} \e^{\i H_2 \Delta t} \bb{u}^{0,1}(t_1)
 = \e^{-p} \bb{u}^1, \quad p\ge 0.
\]
Therefore, the assertion holds for $m=0$.

(2) Assuming the validity of the statement for some positive integer $m=k$, i.e.,
\[\bb{v}^{k+1}(p) = \e^{-p} \bb{u}^{k+1}, \quad p \ge 0,\]
we need to demonstrate that the statement holds for $m = k+1$ with $t\in [t_{k+1},t_{k+2}]$. In this case, Eq.~\eqref{split1} can be reformulated as
\[
\begin{cases}
 \frac{\d}{\d s} \tilde{\bb{v}}^{k+1,1}(s,p) =  - \Lambda \partial_p \tilde{\bb{v}}^{k+1,1}, \quad s \in [0, \Delta t],\\
 \tilde{\bb{v}}^{k+1,1}(0,p) = Q^\dag \bb{v}^{k+1}(p).
 \end{cases}
\]
The solution is
\[ \tilde{\bb{v}}_i^{k+1,1}(s,p) = \tilde{\bb{v}}_i^{k+1,1}(0,p-\lambda_i s) = [\tilde{\bb{v}}^{k+1,1}(0,p-\lambda_i s)]_i
= [Q^\dag \bb{v}^{k+1}(p-\lambda_i s)]_i, \quad i\ge 1,\]
where $[\bb{v}]_i = \bb{v}_i$ represents the $i$-th entry of $\bb{v}$. If $p \ge 0$ and $\lambda_i \ge 0$, we have $p-\lambda_i s\ge 0$. Based on the assumption, we obtain
\[Q^\dag \bb{v}^{k+1}(p-\lambda_i s) = Q^\dag \e^{-(p-\lambda_i s) } \bb{u}^{k+1} = \e^{-p} \e^{\lambda_i s}Q^\dag \bb{u}^{k+1}, \quad p\ge 0.\]
From \eqref{ODEStep1}, we derive
\[Q^\dag \bb{u}^{k+1,1}(t,p) = \e^{\Lambda s} Q^\dag \bb{u}^{k+1} \quad \text{or} \quad
Q^\dag \bb{u}^{k+1} = \e^{-\Lambda s} Q^\dag \bb{u}^{k+1,1}(t,p) ,\]
which leads to
\[Q^\dag \bb{v}^{k+1}(p-\lambda_i s) =  \e^{-p} \e^{\lambda_i s}\e^{-\Lambda s} Q^\dag \bb{u}^{k+1,1}(t,p) , \quad p\ge 0\]
and
\[[Q^\dag \bb{v}^{k+1}(p-\lambda_i s)]_i =  \e^{-p} [Q^\dag \bb{u}^{k+1,1}(t,p)]_i , \quad p\ge 0.\]
Consequently,
\[\tilde{\bb{v}}^{k+1,1}(s,p) = \e^{-p} Q^\dag \bb{u}^{k+1,1}(t,p) , \quad p\ge 0,\]
which means
\[\bb{v}^{k+1,1}(t,p) = \e^{-p} \bb{u}^{k+1,1}(t), \quad p\ge 0.\]
By definition, the second step gives
\[\bb{v}^{k+1,2}(t,p) = \e^{\i H_2 s}\bb{v}^{k+1,1}(t_{k+2},p)
=   \e^{-p} \e^{\i H_2 s} \bb{u}^{k+1,1}(t_{k+2})
= \e^{-p}  \bb{u}^{k+1,2}(t), \quad p\ge 0\]
and
\[\bb{v}^{k+2}(p) = \bb{v}^{k+1,2}(t_{k+2},p) = \e^{-p}  \bb{u}^{k+1,2}(t_{k+2}) = \e^{-p} \bb{u}^{k+2}, \quad p\ge 0.\]
This concludes the proof.
\end{proof}

 \begin{remark}
The statement can be adjusted to accommodate scenarios where the real part matrix $H_1$ possesses positive eigenvalues. In fact, in the first-order time-splitting approximation, the true solution is obtained in the range $p \in [\lambda_+ \Delta t, \infty)$ for the first step, where $\Delta t$ is the time step. Subsequently, the next step covers the range $p \in [2\lambda_+ \Delta t, \infty)$. Therefore, the final range is $p \in [\lambda_+ t, \infty)$, where $t = N_t \Delta t$.
 \end{remark}

We are in a position to derive the corresponding result in Lemma \ref{lem:Pr} in the time-splitting case.

\begin{theorem} \label{thm:Pr}
Suppose that the real part matrix $H_1$ is negative semi-discrete. Let $\bb{v}(t,p)$ be the solution to the time-splitting scheme. Then the probability on the right-hand side of \eqref{Pr} is
\[\text{P}_{\text{r}}(t,p\ge 0) = \frac{\int_0^{\infty} \|\bb{v}(t,p)\|^2 \d p}{\int_{-\infty} ^{\infty} \|\bb{v}(t,p)\|^2 \d p}
=\frac12 \Big(\frac{\|\bb{u}(t)\|}{\| \bb{u}(0) \|}\Big)^2.\]
\end{theorem}

\begin{proof}
(1) It is simple to show that the solutions of \eqref{split1} and \eqref{split2} preserve the $L^2$ norm with respect to $p$ as shown in the proof of Lemma \ref{lem:Pr} and to obtain
\begin{align}
\int_{-\infty}^{\infty} \|\bb{v}(t,p)\|^2 \d p
=  \int_{-\infty}^{\infty} \e^{-2 |p|}  \d p \cdot \| \bb{u}_0 \|^2 = \| \bb{u}_0 \|^2. \label{conserv0}
\end{align}

Suppose that $\lambda_i\le 0$ for $i=1,\cdots,n$ and consider the time interval $[t_m, t_{m+1}]$.
\begin{itemize}
  \item According to Theorem \ref{thm:validility}, the solution of \eqref{split1} in the first step satisfies $\bb{v}^{m,1}(t,p) = \e^{-p} \bb{u}^{m,1}(t)$ for $p>0$, where $\bb{u}^{m,1}(t)$ is the solution of \eqref{ODEStep1}, resulting in
\[\int_0^{\infty} \|\bb{v}^{m,1}(t,p)\|^2 \d p = \int_0^{\infty} \e^{-2p} \d p \|\bb{u}^{m,1}(t)\|^2 = \frac{1}{2} \|\bb{u}^{m,1}(t)\|^2
, \qquad t \in [t_m, t_{m+1}].\]
  \item For the second step, the unitary dynamics of \eqref{split2} and \eqref{ODEStep2} implies
  \[\|\bb{v}^{m,2}(t,p)\| = \|\bb{v}^{m,1}(t_{m+1},p)\|, \qquad
  \|\bb{u}^{m,2}(t)\| = \|\bb{u}^{m,1}(t_{m+1})\|, \qquad t \in [t_m, t_{m+1}],\]
  yielding
\[\int_0^{\infty} \|\bb{v}^{m,2}(t,p)\|^2 \d p = \int_0^{\infty} \|\bb{v}^{m,1}(t_{m+1},p)\|^2 \d p
= \frac{1}{2} \|\bb{u}^{m,2}(t)\|^2, \qquad t \in [t_m,t_{m+1}],\]
or
\[\int_0^{\infty} \|\bb{v}^{m+1}(p)\|^2 \d p = \frac{1}{2} \|\bb{u}^{m+1}\|^2.\]
\end{itemize}
By repeating the above steps, we finally obtain
\[\int_0^{\infty} \|\bb{v}(t,p)\|^2 \d p = \frac{1}{2} \|\bb{u}(t)\|^2, \]
where $t = N_t \Delta t$. The proof is completed by combining the above equation with \eqref{conserv0}.
\end{proof}

\subsection{Finite difference discretization for the heat equation form}

In a recent study by Sato et al. \cite{Sato24Circuit}, a novel approach was introduced for the construction of scalable quantum circuits tailored specifically for wave or Schr\"odinger-type partial differential equations (PDEs), where the Bell basis is employed to diagonalize each term of the Hamiltonian. This methodology has been further extended  in \cite{HuJin24SchrCircuit} to develop quantum circuits for general PDEs, which may not strictly adhere to unitary dynamics, through the application of the Schr\"odingerization technique.

An alternative approach to diagonalizing the discrete Laplacian involves utilizing the Fourier basis. In the subsequent section, we aim to compare these two diagonalization methods: one in the Bell basis and the other in the Fourier basis, in terms of the quantum circuits. Following the approach in \cite{HuJin24SchrCircuit},  we continue to introduce the method utilizing shift operators, which can be diagonalized in the Fourier basis as shown in \cite{Schalker22qBoltzmann}. The resulting diagonal unitary matrix is then evolved using the algorithm described in \cite{Kassal2008Diagonal}, treating it as a black box, as previously done.

In the sequel, we assume the homogeneous Dirichlet boundary conditions for the heat equation form \eqref{FPS}.

\subsubsection{Representation of finite difference operators}

Given a one-dimensional domain $\Omega = [0,L]$, which is uniformly divided into $M = 2^{n_x}$ intervals of length $h = L/M$, one can express a discrete function $u_i$ defined on $x_i = ih$ with $i=0,1,\cdots,M-1$ as $\ket{u} = \sum_{j=0}^{M-1} u_j \ket{j}$ or $\bb{u} = [u_0,\cdots,u_{M-1}]^T$.

An increment operation $S^+$ or a decrement operator $S^-$ takes a quantum state $\ket{j}$ to the state $\ket{j+1}$ or $\ket{j-1}$, namely, \[S^+ \ket{j} = \ket{j+1},  \qquad  S^-\ket{j} = \ket{j-1}, \qquad j = 1,2,\cdots, M-1.\]
These two shift operators are $M \times M$ matrices and can be written as
\[S^+ = \sum\limits_{j=1}^{M-1} \ket{j-1}\bra{j},  \qquad S^- = \sum\limits_{j=1}^{M-1} \ket{j}\bra{j-1}. \]

We will now examine the construction of shift operators as outlined in \cite{Schalker22qBoltzmann}. Consider an integer $k$ with $0 \le k \le 2^{{n_x}-1}$ and its binary representation denoted as $k = (k_0, k_1, \cdots , k_{n-1})$ with the most significant bit on the left, where each $k_i \in \{0,1\}$ for $i = 0,1,\ldots,{n_x}-1$ and $ k = k_0 2^{{n_x}-1} + \cdots  + k_{n_x-1} 2^0$.
We label the qubits from top to bottom as $0,1,\cdots,{n_x}-1$. Denote by $P(\theta)$ the single qubit phase shift gate,
\[P(\theta) = \begin{bmatrix} 1 & 0 \\ 0 & \e^{\i \theta} \end{bmatrix}, \quad
P(\theta) \ket{0} = \ket{0}, \quad P(\theta) \ket{1} = \e^{\i \theta} \ket{1},\]
where $\ket{0}$ and $\ket{1}$ are the basic quantum states for one qubit. Let $\theta_s = \frac{2\pi}{M} 2^{n_x-1-s}$, where $s$ corresponds to the qubit number. It can be observed that
\[\prod_{s = 0}^{n_x-1} P(\theta_s) \ket{k_s} =  \e^{ \frac{\i 2\pi}{M} (2^{n_x-1 - s_1} + \cdots  +  2^{n_x-1 - s_l})} \ket{ k_0, k_1, \cdots, k_{n-1}},\]
where $s_1,\cdots,s_l$ are the indices for which $k_s = 1$ and $\prod$ stands for the tensor product. Apparently,
\[2^{n_x-1 - s_1} + \cdots  +  2^{n_x-1 - s_l} = k_{s_1} 2^{n_x-1 - s_1} + \cdots  +  k_{s_l}2^{n_x-1 - s_l} = k_0 2^{{n_x}-1} + \cdots  + k_{n_x-1} 2^0 = k,\]
leading to
\begin{equation}\label{Upact}
U_{P,+} \ket{k} := \Big( \prod_{s = 0}^{{n_x}-1} P(\theta_s) \Big) \ket{k} =  \e^{ \frac{\i 2\pi k}{M} } \ket{k}.
\end{equation}

\begin{lemma} \label{lem:incdec}
The shift operators $S^+$ and $S^-$ satisfy
\[S^+ = \mathcal{F}^\dag U_{P,+} \mathcal{F}, \qquad S^- = \mathcal{F}^\dag U_{P,-} \mathcal{F},\]
where $\mathcal{F}$ stands for the quantum Fourier transform,
\[U_{P,+} = \prod_{s = 0}^{{n_x}-1} P(\theta_s) , \qquad U_{P,-} = \prod_{s = 0}^{{n_x}-1} P(-\theta_s),  \qquad \theta_s = \frac{2\pi}{N} 2^{n_x-1-s},\]
where $s$ is the qubit number and we label the qubits from top to bottom as $0,1,\cdots,{n_x}-1$.
\end{lemma}
\begin{proof}
By definition,
\[\mathcal{F} \ket{j} = \frac{1}{\sqrt{M}}\sum\limits_{k=0}^{M-1} \e^{\i 2 \pi kj/M} \ket{k}.\]
According to the relation in \eqref{Upact}, one has
\[U_{P,+}\mathcal{F} \ket{j} = \frac{1}{\sqrt{M}}\sum\limits_{k=0}^{M-1} \e^{\i 2 \pi kj/M} \e^{ \frac{\i 2\pi k}{M} } \ket{k}
= \frac{1}{\sqrt{M}}\sum\limits_{k=0}^{M-1} \e^{\i 2 \pi k(j+1)/M} \ket{k} = \mathcal{F} \ket{j+1}. \]
It directly follows that
\[\mathcal{F}^\dag U_{P,+}\mathcal{F} \ket{j} = \ket{j+1},\]
which is exactly the definition of $S^+$.
The proof is completed by noting that $S^- = (S^+)^\dag$.
\end{proof}


The forward and backward difference operators $D^+$ and $D^-$ can be expressed as
\[(D^+\bb{u})_j = \frac{u_{j+1}-u_j}{h}, \qquad (D^-\bb{u})_j = \frac{u_j-u_{j-1}}{h}, \quad j = 0,1,\cdots, M-1, \]
where $u_M$ and $u_{-1}$ are determined from the boundary conditions (BCs). For instance,
\[u_M := \begin{cases}
0 \quad & \mbox{for Dirichlet BC} \\
u_{M-1} \quad & \mbox{for Neumann BC} \\
u_0 \quad & \mbox{for periodic BC}
\end{cases}, \qquad
u_{-1} := \begin{cases}
0 \quad & \mbox{for Dirichlet BC} \\
u_0 \quad & \mbox{for Neumann BC} \\
u_{M-1} \quad & \mbox{for periodic BC}
\end{cases}.\]
The discrete Laplacian operator $D^\Delta$ is then defined as
\[(D^\Delta \bb{u})_j =  (D^+D^-\bb{u})_j = \frac{u_{j+1} - 2 u_j + u_{j-1}}{h^2}, \qquad j = 0,1,\cdots, M-1.\]

A straightforward manipulation gives (cf. \cite{Sato24Circuit}):
\[D_D^\Delta = \frac{S^- + S^+ - 2 I}{h^2},\]
where the Dirichlet boundary condition is imposed. By Lemma \ref{lem:incdec}, we also have
\[D_D^\Delta = \frac{1}{h^2} \mathcal{F}^\dag ( U_{P,-} + U_{P,+} - 2 I )\mathcal{F}.\]

\subsubsection{Quantum simulation for finite difference discretizations}

For simplicity, we change the variable $\psi$ by $u$ and consider the heat equation
\[\partial_t u = a \Delta u - U(x)u, \qquad a = \sigma,\]
where the Dirichlet boundary condition is imposed. We focus only on the one-dimensional case since the extension to higher dimensions is straightforward. The numbers of qubits for $x$ and $p$ registers are denoted by $n_x$ and $n_p$, respectively.

Through the application of the time-splitting technique, the problem reduces to $\partial_t u = a \Delta u$ with $t\in [0,\Delta t]$, where $\Delta t$ is the time step, for which case the semi-discrete problem is
\[\frac{\d \bb{u}}{\d t} = D_D^\Delta \bb{u}, \qquad  \bb{u}(0) = \bb{u}_0. \]
The Schr\"odingeried system associated with \eqref{generalSchr} is
\[\frac{\d}{\d t} \tilde{\bb{w}}(t) = -\i ( D_D^\Delta \otimes D_\mu ) \tilde{\bb{w}}.\]
By the change of variables $\bb{\psi} = (\mathcal{F}^\dag \otimes I_p) \tilde{\bb{w}}$, one has
\[\frac{\d}{\d t} \bb{\psi}(t) = -\i\frac{1}{h^2} ( U_{P,-} + U_{P,+} - 2 I_x ) \otimes D_\mu  \bb{\psi}
\equiv -\i (H_1 \otimes D_\mu ) \bb{\psi} \equiv -\i V \bb{\psi}.\]
According to \eqref{Upact}, $U_{P,+}$ is a diagonal matrix with the $k$-th entry on the diagonal given by
$(U_{P,+})_{k,k} = \e^{ \frac{\i 2\pi k}{M} }$, where $k=0,1,\cdots,M-1$. This implies that
\[(H_1)_{k,k} = \frac{1}{h^2}  (\e^{ \frac{\i 2\pi k}{M} } +  \e^{ -\frac{\i 2\pi k}{M} } - 2) = - \frac{4}{h^2} \sin ^2 \frac{ \pi k}{M} =:-d_k \le 0, \]
which are the eigenvalues of $D_D^\Delta$.

The quantum circuit is similar to the one given in Fig.~\ref{fig:circuitheat}, where the diagonal unitary is treated as a black box, the problem simplifies to the time evolution of a diagonal unitary matrix $\e^{-\mathrm{i} V \Delta t}$, a task efficiently executable with $\mathcal{O}(n_x+n_p)$ gates when neglecting the cost of evaluation of $V$, as demonstrated by the algorithm outlined in \cite{Kassal2008Diagonal}.

In the following, we elaborate on this algorithm.
For brevity, we denote $\ket{\bb{x}} = \ket{x}\ket{p}$, where $0\le x< M_x-1$ and $0\le p<M_p-1$ are integers with $M_x = 2^{n_x}$ and $M_p = 2^{n_p}$. Accordingly, we write $V(\bb{x}) = d_x \mu_p$, where $\mu_p$ is the diagonal entry of $D_\mu$ for $p = 0,\cdots, M_p-1$. The application of the diagonal unitary is equivalent to the following map:
\[\ket{\bb{x}} \mapsto \e^{-\i V(\bb{x}) \Delta t} \ket{\bb{x}},\]
since $\bb{\psi}$ is the superposition of ``position states'' $ \ket{\bb{x}}$.
\begin{itemize}
  \item Without loss of generality, we can assume that $V(\bb{x}) \ge 0$ for all $\bb{x}$. Indeed, let $c = V_{\min}$. The evolution can be written as  $\e^{-\i (V(\bb{x}) +  |c| - |c|) }  = \e^{\i |c|} \e^{-\i (V(\bb{x}) + |c| ) }$. The global phase $\e^{\i |c|}$ has no impact on the time evolution.
  \item Let $n = n_x + n_p$. We can further assume that $ 2\pi V(\bb{x}) \le  M-1 $ by choosing $n_x$ and $n_p$, where $M = 2^n = M_x M_p$. In this case, $0\le \frac{ 2\pi V(\bb{x})}{M} \le 1- \frac{1}{2^{n-1}}$, meaning that $\frac{2\pi V(\bb{x})}{M}$ can be represented by an $n$-bit string as $(0.k_1 k_2 \cdots k_n)$ with given precision $\epsilon$ when $n$ scales as $\mathcal{O}(\log \epsilon^{-1} )$, where $(0.k_1 k_2 \cdots k_n) = \sum\limits_{i=1}^n \frac{k_i}{2^i} $. For this reason, we choose the time step such that
$\Delta t = 2\pi /M = \mathcal{O}(\epsilon)$, implying that $\frac{ 2\pi V(\bb{x})}{M} \approx \frac{ 2\pi q_{\bb{x} }}{M}$ has additive precision $\mathcal{O}(\epsilon)$, where $q_{\bb{x}} = [V(\bb{x})]$ is the integer part of $V(\bb{x})$. This means that we can replace $V(\bb{x})$ by $q_{\bb{x}}$ when the time step satisfies the prescribed condition.
\end{itemize}

The algorithm requires $n$ ancilla qubits. The quantum computer is initialized in the state $\ket{\psi} \otimes \ket{1}_n = \sum_{\bb{x}} \psi(\bb{x}) \ket{\bb{x}}\otimes \ket{1}_n $, where $\ket{1}_n$ in the ancilla register represents the state $\ket{0\cdots 001}$ in $n$ qubits. Apply the QFT to the ancilla register produces
\begin{equation}\label{ancy}
(I \otimes \mathcal{F}) (\ket{\bb{x}} \otimes \ket{1}_n) = \ket{\bb{x}} \otimes \frac{1}{\sqrt{M}}\sum\limits_{y=0}^{M-1} \e^{\i 2 \pi y/M}\ket{y}.
\end{equation}
The diagonal unitary is then applied by phase kickback described as follows.

\begin{itemize}
  \item The basic idea is to establish the following map:
  \begin{equation}\label{Vact}
  \mathcal{V} \ket{\bb{x},y}_{\oplus} = \e^{-2\pi \i q_{\bb{x}}/M} \ket{\bb{x},y}, \qquad q_{\bb{x}} = [V(\bb{x})],
  \end{equation}
  since it produces
   \begin{align*}
  \mathcal{V}\Big(\ket{\bb{x}} \otimes \frac{1}{\sqrt{M}}\sum\limits_{y=0}^{M-1} \e^{\i 2 \pi y/M}\ket{y}\Big)
  & = \e^{-2\pi \i q_{\bb{x}} / M} \ket{\bb{x}} \otimes \frac{1}{\sqrt{M}}\sum\limits_{y=0}^{M-1} \e^{\i 2 \pi y/M}\ket{y}\\
  & \approx \e^{-2\pi \i V({\bb{x}}) / M} \ket{\bb{x}} \otimes \frac{1}{\sqrt{M}}\sum\limits_{y=0}^{M-1} \e^{\i 2 \pi y/M}\ket{y} \\
  & = \e^{-\i V(\bb{x}) \Delta t} \ket{\bb{x}} \otimes \frac{1}{\sqrt{M}}\sum\limits_{y=0}^{M-1} \e^{\i 2 \pi y/M}\ket{y}.
  \end{align*}

  \begin{figure}[!htb]
  \centering
  \includegraphics[scale=0.2]{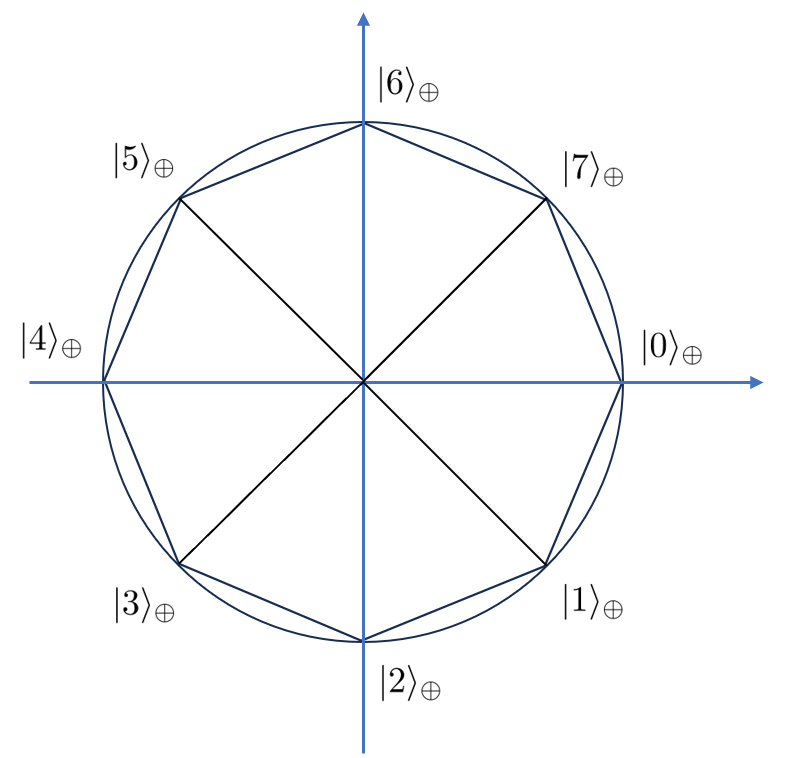}\\
  \caption{The eigenstates of addition modulo $2^n$ in $\mathbb{C}^2$}\label{fig:cyclic}
\end{figure}

  \item The basis state $\ket{y}$ of the ancilla register can be mapped to a vector $\ket{y}_{\oplus} = \e^{-2\pi \i y/M}$ in $\mathbb{C}^2$. As shown in Fig.~\ref{fig:cyclic}, when $M = 2^3 = 8$, these vectors correspond to eight equally spaced points on the unit circle in the complex plane, arranged clockwise.  Let $\theta = 2\pi/M$ and $q\ge 0$ be an integer. For any vector $\ket{y}_{\oplus}$ (where $y = 0,\cdots, M-1$), the action of $\mathcal{V}$ in \eqref{Vact} can be treated as the operation of rotating $\ket{y}_{\oplus}$ clockwise by $q \theta$.

  \item  This rotation precisely corresponds to the addition modulo $2^n = M$, i.e., $\ket{y \oplus q}$, where $y \oplus q$ denotes addition modulo $2^n$. Take Fig.~\ref{fig:cyclic} as an example. The expression
      $\ket{4 \oplus 5} = \ket{1}$ signifies the clockwise rotation of $\ket{4}_{\oplus}$ to $\ket{1}_{\oplus}$, with a rotation angle of $5 \theta$. The geometric meaning implies
      \[\ket{y\oplus q} = \e^{-2\pi \i q/M} \ket{y}. \]
  Because of this, the definition in \eqref{Vact} can be modified as
  \[\mathcal{V} \ket{\bb{x},y} = \ket{\bb{x},y\oplus q_{\bb{x}} }, \qquad q_{\bb{x}} = [V(\bb{x})].\]
\end{itemize}

The operator $\mathcal{V}$ corresponds to the circuit for classical computation using reversible gates. In fact,  any irreversible classical gate $x \mapsto f(x)$ can be made into
a reversible classical gate $(x,y) \mapsto (x, y\oplus f(x))$ \cite{NC2010Quantum,Lin2022Notes}.  Since all reversible single-bit and two-bit classical gates can be realized using single-qubit and two-qubit quantum gates, the reversible mapping can be represented as a unitary operator $\ket{x,y} \mapsto \ket{x, y\oplus f(x)}$ on a quantum computer. This proves that a quantum computer is at least as powerful as classical computers.

Based on the preceding discussion, in computing the runtime of the circuit depicted in Fig.~\ref{fig:circuitheat}, we have actually neglected the classical operations for $V$, as the classical method also undergoes this process. This suggests that the quantum algorithm utilizing the time-splitting technique is more potent than its classical counterpart, since it requires exponentially more operations to compute the Fourier transform on a classical computer than it does to implement the quantum Fourier transform on a quantum computer.

However, implementing reversible computing requires detailed technical considerations for realizing reversible single-bit and two-bit classical gates using single-qubit and two-qubit quantum gates. This inevitably leads to a significant overhead in terms of the number of ancilla qubits and circuit depth compared to the explicit quantum circuits presented in \cite{HuJin24SchrCircuit}. In fact, as illustrated in Figures 1-4 of \cite{HuJin24SchrCircuit}, only two fixed parameters, $\gamma_0$ and $\tau$, are utilized in the circuits. Hence, the classical computation involved in the circuit can be disregarded, which is very different from the reversible computation. In summary, diagonalization in the Bell basis proves to be much more efficient than diagonalization in the Fourier basis in general cases.

\section{Conclusion} \label{sec:conclusion}

 We use the Schr\"odingerization approach introduced in \cite{JLY22SchrShort, JLY22SchrLong} for quantum simulation of the Fokker-Planck equation. While the semi-discretization is not necessarily a Hermitian Hamiltonian system, which is most natural for quantum simulation, the Schr\"odingerization approach makes it so in a simple fashion.

 We provide the implementation details for the conservation and heat equation forms of the Fokker-Planck equation.
For the conservation form, we investigate two different approaches to implementing the Schr\"odingerization procedure and point out that
the exponential change of variables in \cite{ALL2023LCH} has an equivalent impact on projecting onto the solution vector.  Regarding the heat equation form, we concentrate on the quantum simulation procedure using the time-splitting technique. We examine the relationship between operator splitting in the Schr\"odingerization method and its direct application to the original problem, and conduct a comparative study of diagonalization in the Bell and Fourier bases for finite difference discretizations.


\section*{Acknowledgements}
SJ and NL are supported by NSFC grant No. 12341104, the Shanghai Jiao Tong University 2030 Initiative and the Fundamental Research Funds for the Central Universities. SJ was also partially supported by the NSFC grants Nos. 12031013, the Shanghai Municipal Science and
Technology Major Project (2021SHZDZX0102), and the Innovation Program of Shanghai Municipal Education Commission (No. 2021-01-07-00-02-E00087). NL also acknowledges funding from the Science and Technology Program of Shanghai, China (21JC1402900).
YY is supported by the National Science Foundation for Young Scientists of China (No. 12301561).


\newcommand{\etalchar}[1]{$^{#1}$}

\end{document}